\documentclass[a4paper,aps,prd,floatfix,reprint,twoside,twocolumn,preprintnumbers,showkeys,showpacs,final]{revtex4-2}
\usepackage[utf8]{inputenc}
\usepackage{amsmath}
\usepackage{amssymb}
\usepackage{color}
\usepackage{natbib}
\usepackage{graphicx}
\usepackage{xr-hyper}
\usepackage{hyperref} % links with BibTeX
\usepackage{enumitem}
\usepackage[caption=false]{subfig}
\usepackage{stackengine}
\usepackage{afterpage}
\usepackage{float}
\usepackage{array}
\usepackage{soul}
\DeclareMathOperator{\Tr}{Tr}
\newcommand{\del}{\partial}

\graphicspath{{./Vector/}{./}}

\definecolor{lightblue}{rgb}{0.70,0.75,1}
\sethlcolor{lightblue}

\begin{document}
\preprint{ADP-19-29/T1109}

\title{Visualisation of Centre Vortex Structure}
\author{James C. Biddle}
\author{Waseem Kamleh}
\author{Derek B. Leinweber}
\affiliation{Centre for the Subatomic Structure of Matter, Department of Physics, The University of Adelaide, SA 5005, Australia}

\begin{abstract}
  The centre vortex structure of the $SU(3)$ gauge field vacuum is explored through the
  use of novel visualisation techniques. The lattice is partitioned into 3D time slices,
  and vortices are identified by locating plaquettes with nontrivial centre phases.
  Vortices are illustrated by rendering vortex lines that pierce these nontrivial
  plaquettes. Nontrivial plaquettes with one dimension in the suppressed time direction
  are rendered by identifying the visible spatial link. These visualisations highlight the
  frequent presence of singular points and reveal an important role for branching points
  in $SU(3)$ gauge theory in creating high topological charge density regimes.
  Visualisations of the topological charge density are presented, and an investigation
  into the correlation between vortex structures and topological charge density is
  conducted. The results provide new insight into the mechanisms by which centre vortices
  generate nontrivial gauge field topology. This work demonstrates the utility of
  visualisations in conducting centre vortex studies, presenting new avenues with which to
  investigate this perspective of the QCD vacuum.
\end{abstract}

\pacs{12.38.Gc,12.38.Aw,14.70.Dj}
\keywords{Data Visualisation; Centre Vortices; Lattice QCD; Topological Charge}

\maketitle

\section{Introduction}\label{sec:Intro}

In recent years the centre vortex perspective of the QCD vacuum~\cite{'tHooft:1977hy,'tHooft:1979uj} has emerged as the most fundamental aspect of QCD vacuum structure, simultaneously governing the properties of confinement and dynamical chiral symmetry breaking in quantum chromodynamics (QCD). Centre vortices have been shown to give rise to mass splitting in the low-lying hadron spectrum~\cite{Trewartha:2017ive,Trewartha:2015nna,OMalley:2011aa}, a linear static quark potential~\cite{Cais:2008za, Langfeld:2003ev, Trewartha:2015ida,Greensite:2003bk,DelDebbio:1998luz}, appropriate Casimir scaling~\cite{Faber:1997rp}, appropriate behaviour of the quark propagator~\cite{Bowman:2008qd,Trewartha:2015nna} and infrared enhancement of the gluon propagator~\cite{Bowman:2010zr, Biddle:2018dtc, Langfeld:2001cz, Quandt:2010yq}. These results all support the theory that centre vortices capture the essence of QCD vacuum structure and contribute significantly to a full understanding of QCD.

Centre vortices naturally give rise to an area-law falloff in the Wilson loop expectation value~\cite{Greensite:2016pfc}, such that
\begin{equation}
\langle W(C) \rangle \propto \exp\left(-\sigma A(C)\right)\, ,
\end{equation}
where $A(C)$ is the minimal area spanned by the Wilson loop. This area-law behaviour is often taken to be an indicator of confinement~\cite{DelDebbio:1998luz,Dosch:1988ha}. The fact that one of the defining features of the vortex model is tied so intimately to the geometry of vortices in the vacuum indicates that visualising these structures may provide valuable insight~\cite{Biddle:2019ctr, Biddle:2019abu}. To this end, we construct visualisations of centre vortices and topological charge density on the lattice. We use these visualisations to investigate the dynamics of the vortex model in an interactive and novel manner.

We begin this work in Sec.~\ref{sec:VortexIdentification} with a description of how centre vortices are identified on Monte-Carlo generated lattice gauge fields. In Sec.~\ref{sec:SOVortices} we then describe in detail our convention for plotting vortices in three dimensional space, and present the first interactive visualisations of centre vortices on the lattice.

In the Supplemental Material\cite{SuppMaterial} these visualisations are presented as interactive 3D models embedded in the document. To interact with these models, it is necessary to open the Supplemental document in Adobe Reader or Adobe Acrobat (requires version 9 or newer). Linux users may install Adobe Acroread version 9.4.1, the last edition to have full 3D support. Note that 3D content must also be enabled for the interactive content to be available, and for proper rendering it is necessary to enable double-sided rendering in the preferences menu. Figures with a corresponding interactive model that can be found in the Supplemental material are marked as \textbf{Interactive} in the caption. Interactive models in the Supplemental Material are also referenced as Fig.~S-xx in the text. To activate the models, simply click on the image. To rotate the model, click and hold the left mouse button and move the mouse. Use the scroll wheel or shift-click to zoom. Some preset views of the model are also provided to highlight areas of interest. To reset the model back to its original orientation and zoom, press the ``home'' icon in the toolbar or change the view to ``Default view''.

As projected centre vortices are inherently two-dimensional objects embedded in four dimensions, we describe the technique used to capture the behaviour of vortices in the fourth dimension in Sec.~\ref{sec:STOVortices}. In Sec.~\ref{sec:TopChargeVis} we present visualisations of topological charge density alongside vortex lines. In Secs.~\ref{sec:SP} and \ref{sec:BP} we describe singular points and branching points; two of the unique vortex structures present in the vacuum. Finally, in Sec.~\ref{sec:Correlation} we investigate the correlation these structures have with topological charge density. This investigation lays the groundwork for the development of further visualisation techniques, and emphasises the importance of centre vortex geometry in a full understanding of the QCD vacuum.

\section{Vortex Identification}\label{sec:VortexIdentification}
To visualise vortices, we first need to outline how they are identified on the lattice. Physical centre vortices are ``thick'' objects, meaning that in four dimensions they form sheets of finite thickness~\cite{Nielsen:1979xu}. In contrast, on the lattice we identify ``thin'' or ``projected'' vortices, known as P-vortices. These P-vortices have  infinitesimal thickness. Whilst not physical, P-vortices have been shown to be highly correlated to the location of thick vortices, indicating that it is appropriate to concern ourselves with the behaviour of P-vortices in understanding vortex structure on the lattice~\cite{DelDebbio:1998luz}.

We perform our calculations on 100 $20^3\times 40$ $SU(3)$ gauge field configurations of lattice spacing $a=0.125\,\text{fm}$. To identify P-vortices, we first gauge-transform each configuration to maximal centre gauge. This is performed by creating a gauge transformation $\Omega(x)$ that maximises the functional~\cite{Langfeld:2003ev}
\begin{equation}
\label{eq:MCGfunctional}
R=\frac{1}{V\,N_{\text{dim}}\,n_c^2}\sum_{x,\mu}|\text{Tr}\,U_{\mu}^\Omega(x)|^2\, .
\end{equation}
Maximising Eq.~\eqref{eq:MCGfunctional} serves to bring every link $U_\mu(x)$ as close as possible to the centre of $SU(3)$, which consists of the three elements
\begin{equation}
\mathbb{Z}_3 = \big\lbrace \exp\left(\frac{\pm 2\pi i}{3} \right)I,\, I\big\rbrace\, .
\end{equation}
Once the configuration is fixed to maximal centre gauge, we then project $U_\mu(x)$ onto the nearest centre element to obtain our vortex-only configurations, $Z_\mu(x)$. It is these vortex-only configurations that we shall be working with for constructing our visualisations.

Once we have obtained our vortex-only configurations, it is simple to identify centre vortices. As we are concerned with P-vortices, it is sufficient to calculate the smallest $1\times 1$ Wilson loop on the lattice. In the $\mu,\,\nu$ plane we denote the plaquette by $P_{\mu\nu}(x)$. As the centre of $SU(3)$ is closed and $P_{\mu\nu}$ is the product of four centre elements, $P_{\mu\nu}(x)$ is itself a centre element. Noting that vortices live on the dual lattice, the centre flux associated with a plaquette is given by~\cite{Engelhardt:2003wm,Spengler:2018dxt}
\begin{equation}
  \label{eq:dualplaq}
  P_{\mu\nu}(x) = \exp\left(\frac{\pi i}{3}\,\epsilon_{\mu\nu\kappa\lambda}m_{\kappa\lambda}(\bar{x})\right),
\end{equation}
where $m_{\kappa\lambda}(\bar{x})$ is the (oriented) elementary square anchored at the point $\bar{x} = x + \frac{a}{2}(\hat{\mu} + \hat{\nu} - \hat{\kappa} - \hat{\lambda})$ on the dual lattice, such that it pierces the plaquette. $m_{\kappa\lambda}(\bar{x})$ is antisymmetric under index permutation.

If $P_{\mu\nu}(x) \neq I$ then we say that the plaquette is pierced by a centre vortex of charge $m_{\kappa\lambda}(\bar{x}) = -m_{\lambda\kappa}(\bar{x}) = \pm 1$, otherwise if $P_{\mu\nu}(x)=I$ then $m_{\kappa\lambda}(\bar{x})=0$ and we say it is not pierced by a vortex. Note that the epsilon tensor in Eq.~\ref{eq:dualplaq} removes any ambiguity in the assignment of the vortex phase associated with the ordering of the Lorentz indices. The orientation of vortices is significant to the behaviour of vortex models as a whole, and is discussed in greater detail in Ref.~\cite{Engelhardt:1999xw}.

Now that we have identified P-vortices on the lattice, we can begin to construct 3D visualisations. These visualisations aim to elucidate the properties of vortices, and serve as a guide to explaining how vortex structures give rise to the salient features of QCD.

\section{Spatially-Oriented Vortices}\label{sec:SOVortices}
\subsection{Visualisation Conventions}
As the lattice is a four-dimensional hypercube, we visualise the centre vortices on a set of 3D slices. The choice of dimension to take slices along is irrelevant at low temperature in Euclidean space where our lattice calculations take place, so to maximise the volume of each slice we introduce a coordinate system with the $z$ axis along the long dimension, and take slices along the $t$ axis. This results in $N_t=20$ slices each with dimensions $N_x\times N_y\times N_z=20\times 20 \times 40$. Within each slice we can visualise all vortices associated with an $x-y$, $x-z$ or $y-z$ spatial plaquette by calculating $P_{x\,y}(\bf{x})$, $P_{y\,z}(\bf{x})$ and $P_{z\,x}(\bf{x})$ for all $\bf{x}$ in the slice. These vortices will be referred to as the ``spatially-oriented'' vortices, as they are fixed in time.

As discussed in the previous section, the plaquettes are evaluated on a centre projected configuration, so we can identify the spatial plaquettes with the group of integers modulo $3$ according to the vortex centre charge $m_k \in \lbrace -1,\,0,\,+1\rbrace$, such that $P_{ij} = \exp\left(\frac{2\pi i}{3}\epsilon_{ijk}m_k\right)I$. Hereafter we will refer to a plaquette simply by its centre charge.

\begin{figure}
  \includegraphics[width=\linewidth]{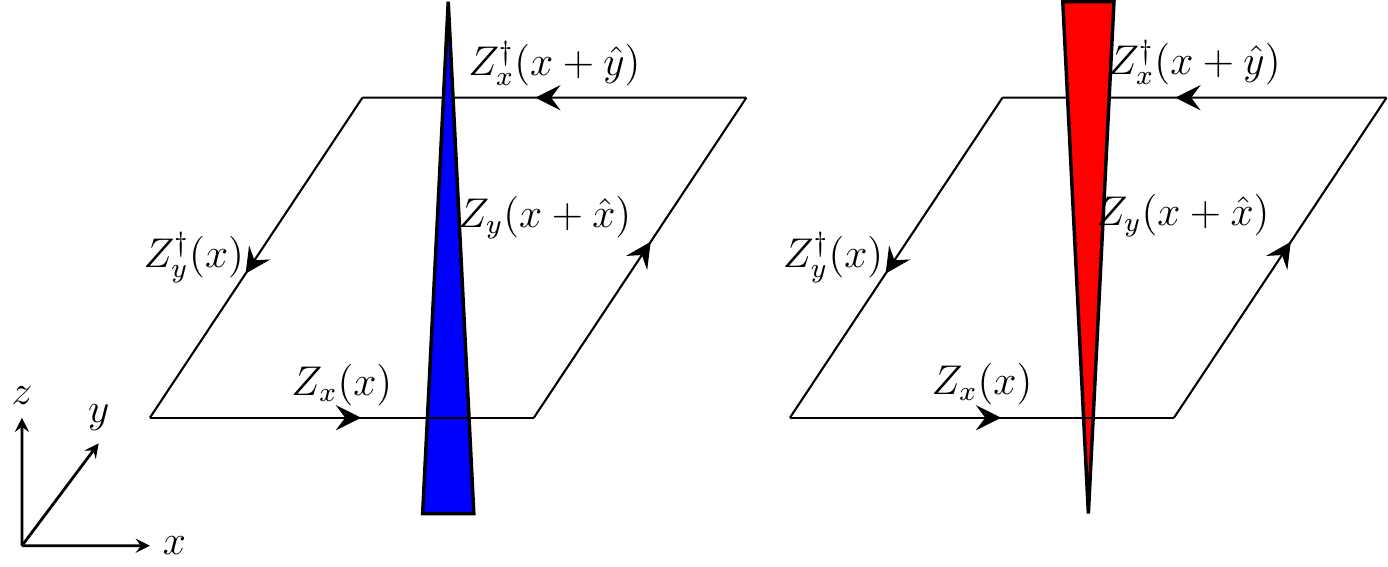}
\caption{\label{fig:SpacialVortices}An example of the plotting convention for vortices located within a 3D time slice. \textit{Left:} A $+1$ vortex in the $+\hat{z}$ direction. \textit{Right:} A $-1$ vortex in the $-\hat{z}$ direction.}
\end{figure}

For a charge $m_k=+1$ vortex, a blue jet is plotted piercing the centre of the plaquette, and for a charge $m_k=-1$ vortex, a red jet is plotted. The direction of the jet is set according to the right-hand rule of the epsilon tensor, such that
\begin{itemize}[leftmargin=*,itemsep=0pt,labelsep=12pt]
\item $P_{x\,y}=\pm 1\implies \pm\hat{z}$ direction.
\item $P_{y\,z}=\pm 1\implies \pm\hat{x}$ direction.
\item $P_{z\,x}=\pm 1\implies \pm\hat{y}$ direction,
\end{itemize}
An example of this plotting convention is shown in Fig.~\ref{fig:SpacialVortices}. As the jet direction $\pm \hat{k}$ is aligned with the sign of the centre charge $m_k = \pm 1$ the vortex lines show the oriented flow of positive unit centre charge ($m = +1$).

Projected centre vortices are surfaces in four-dimensional space-time, analogous
to the centre line of a vortex in fluid dynamics that maps out a surface as it moves through time. Note that, as is conventional, herein "time" simply refers to the fourth spatial dimension on the Euclidean lattice. Similarly, "time evolution" refers to change with respect to the fourth spatial dimension, that is, variation in Euclidean time (not real time). In this way, the visualisations presented here can be simply thought of as a way to interpret the four-dimensional geometry. As the surface cuts through the three-dimensional spatial volume of our visualization, a P-vortex line  is rendered mapping the flow of centre charge.

\subsection{Bianchi Identity}

The centre vortex flux satisfies the Bianchi identity,
\begin{equation}
  \label{eq:bianchi}
  \epsilon_{\mu\nu\kappa\lambda} \, \del_\kappa \, F_{\mu\nu}(x) = 0,
\end{equation}
such that the flux through a spatial cube is
conserved~\cite{Engelhardt:2003wm,Spengler:2018dxt}. This can be easily seen. We start by
noting that the field strength tensor can be related to the plaquette by
\begin{equation}
  \label{eq:fmunu}
  iga^2F_{\mu\nu}(x) = 1 - P_{\mu\nu}(x) + O(a^3)
\end{equation}
where $g$ is the gauge coupling and $a$ is the lattice spacing. Drawing on Eq.~\ref{eq:dualplaq} and noting that $\frac{\del \bar{x}_\alpha}{\del x_\beta} = \delta_{\alpha\beta}$, the Bianchi identity of Eq.~\ref{eq:bianchi} becomes
\begin{equation}
  \epsilon_{\mu\nu\kappa\lambda}\, \epsilon_{\mu\nu\sigma\tau}\, \bar{\del}_\kappa\, m_{\sigma\tau}(\bar{x}) = 0,
\end{equation}
where $\bar{\del}_\kappa = \frac{\del}{\del \bar{x}_\kappa}$. Recalling
$m_{\kappa\lambda}$ is antisymmetric and
\begin{equation}
  \epsilon_{\mu\nu\kappa\lambda}\, \epsilon_{\mu\nu\sigma\tau} = 2\left(\delta_{\kappa\sigma} \delta_{\lambda\tau} - \delta_{\kappa\tau} \delta_{\lambda\sigma} \right),
\end{equation}
one finds
\begin{equation}
  \bar{\del}_\kappa\, m_{\kappa\lambda}(\bar{x}) = 0.
\end{equation}

To make contact with our 3D visualisation of spatial plaquettes where $\mu,\,\nu,\,\kappa \in i,\,j,\,k = 1,\,2,\,3$, we set $\mu = i$, $\nu = j$, $\kappa = k$, $\lambda = 4$ and examine the spatial divergence
\begin{equation}
  \bar{\del}_k\, m_{k 4}(\bar{x}) = \vec{\nabla} \cdot \vec{m}(\bar{x}) = 0,
\end{equation}
with $m_{\kappa 4} = m_k = \left[ \vec{m} \right]_k$ being the spatially-oriented vortex flux piercing the spatial plaquette $P_{ij}(x)$. Recalling the divergence theorem
\begin{equation}
  \int_V d^3r\, \vec{\nabla} \cdot \vec{m}(\vec{r}) = \int_{\del V} d\vec{S} \cdot \vec{m}(\vec{r}) = 0,
\end{equation}
and $m_k = \pm 1$, centre-vortex flux entering a face of a spatial cube $V$ has to leave by another face. In our visualisations, $m_{k} = \pm 1$ is represented by a jet in the $\pm\hat{k}$ direction plotted at $\tilde{x} = x + \frac{a}{2}\left(\hat{i} + \hat{j} - \hat{k} \right)$, such that the above implies a continuous flow of centre vortex flux through a spatial cube.

\begin{figure}[t]
\includegraphics[width=\linewidth]{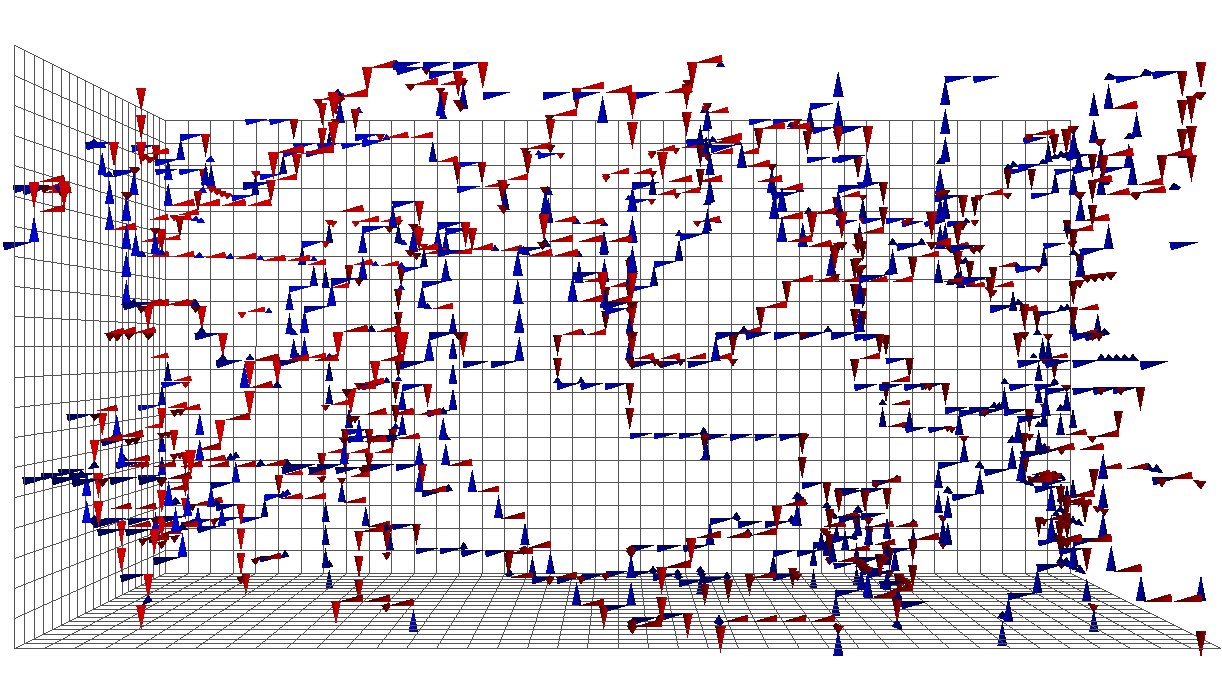}
  \caption{\label{fig:PlaqT01}The $t=1$ slice with all spatially-oriented vortices plotted.  The flow of $m=+1$ centre charge is illustrated by the jets as described in the text. (\textbf{Interactive}. See Sec.~\ref{sec:Intro} for more information about interacting with the 3D models available in the supplementary material.)}
\end{figure}

\begin{figure}[t]
\includegraphics[width=\linewidth]{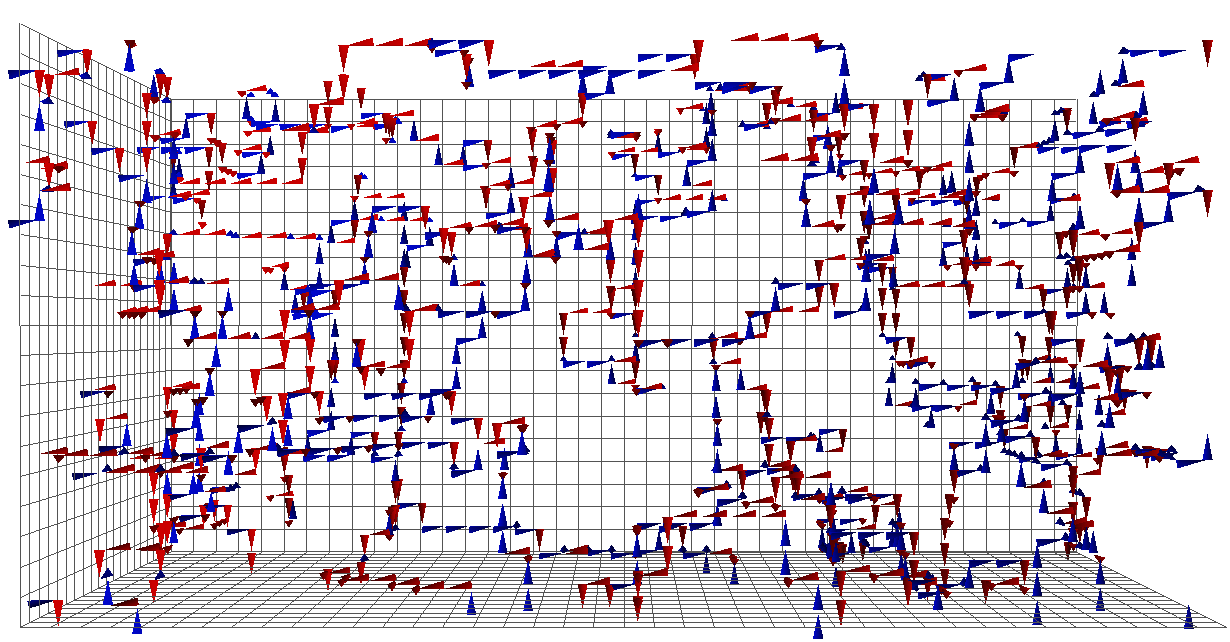}
  \caption{\label{fig:PlaqT02}The $t=2$ slice with all spatially-oriented vortices plotted. Only a small subset of jets are stationary between $t=1$ and $t=2$. Symbols are as in Fig.~\ref{fig:PlaqT01}. (\textbf{Interactive})}
\end{figure}

\begin{figure}[t]
    \subfloat{
	\includegraphics[width=0.3\linewidth]{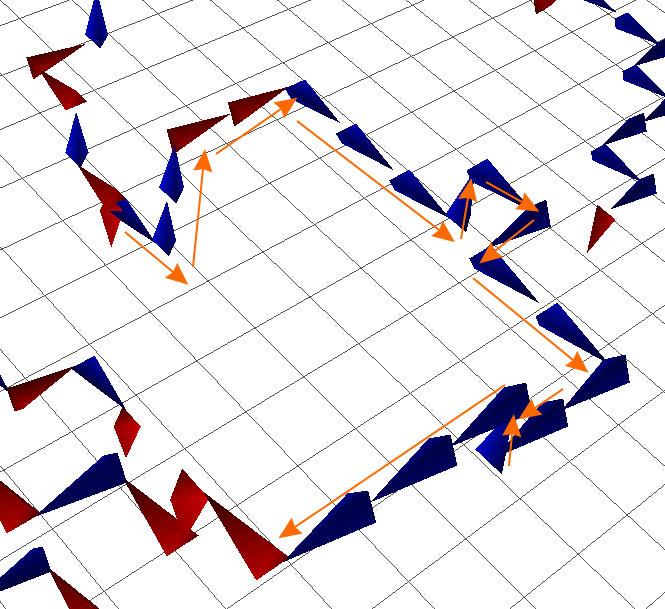}
    }\hfill
    \subfloat{
    \includegraphics[width=0.3\linewidth]{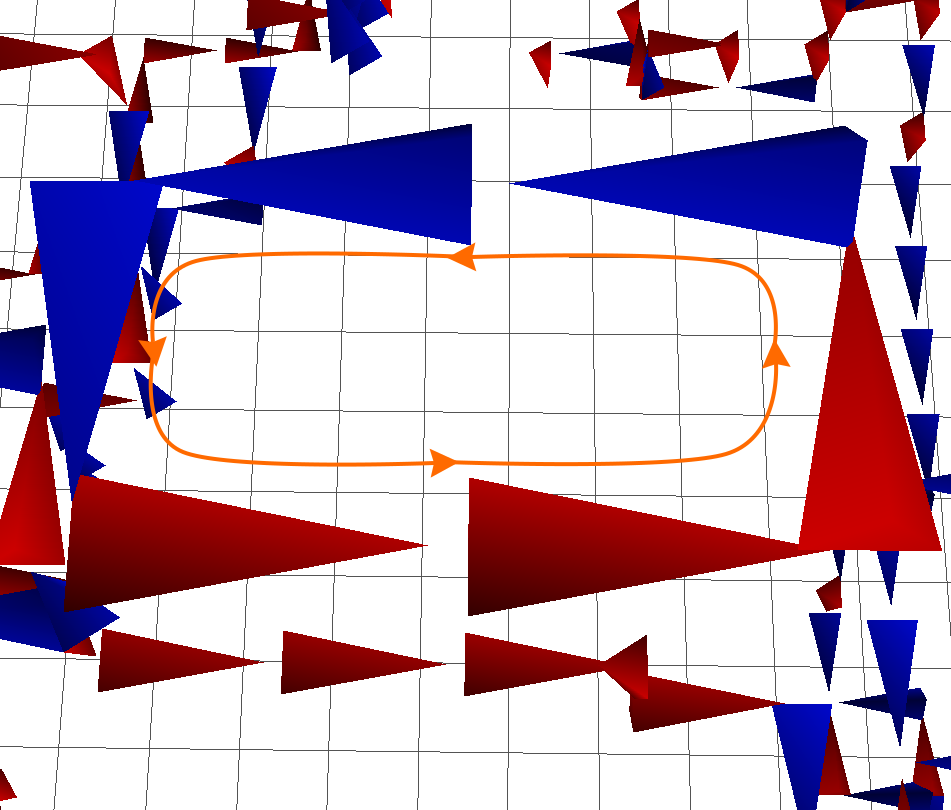}
    }\hfill
    \subfloat{
	\includegraphics[width=0.3\linewidth]{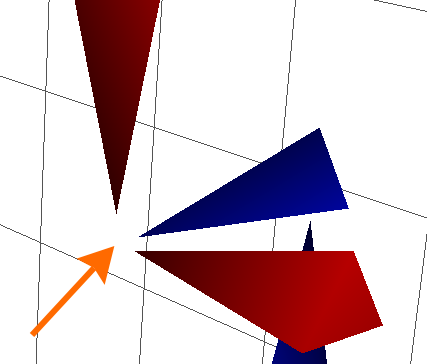}
    }
    \caption{\label{fig:VortexFeatures} \textit{Left:} Vortices form directed continuous
      lines, highlighted with orange arrows in this diagram. Note that because of the
      lattice periodicity, these lines may wrap around to the opposite edge of the
      lattice. \textit{Middle:} Vortices must form closed loops to conserve the vortex
      flux. \textit{Right:} $SU(3)$ vortices are capable of forming monopoles or branching
      points where three or five vortices emerge or converge at a single point.}
\end{figure}

The spatially-oriented vortices for the 3D slices with $t=1,2$ are illustrated in Figs.~\ref{fig:PlaqT01}, \ref{fig:PlaqT02}. At first glance the vortex structure appears highly complex, and it is difficult to identify the significant features. As such, we make use of the 3D models to hone in and isolate the important features present in these slices. We present some of these features in Fig.~\ref{fig:VortexFeatures}. We observe that the vortices do indeed form closed lines (as required by the Bianchi identity), highlighted in the view ``Vortex path'' in Fig.~S-1 and the middle panel of Fig.~\ref{fig:VortexFeatures}.  We also see that the vortex loops tend to be large. This agrees with the determination made of $SU(2)$ vortices in Refs.~\cite{Engelhardt:1999fd,Bertle:1999tw}.

\subsection{Branching/Monopole Points}

The presence of branching/monopole points is of particular interest, as previous studies
have primarily focussed on $SU(2)$ theory which is free from these structures. In $SU(3)$
it is possible to conserve centre flux at the intersection of three or five vortex lines within a
3D slice. An example of a branching/monopole point in our visualisations is shown in the
right panel of Fig.~\ref{fig:VortexFeatures} and the interactive view `Monopole' in
Fig.~S-1.

The ambiguity between monopoles and branching points~\cite{Spengler:2018dxt} arises from the periodicity of the centre phase $z = \exp\left(2\pi i/3 \right)$. By our conventions, each jet denotes the directed flow of $+1$ centre charge. However, because $\exp(2\pi i/3) = \exp(-4\pi i/3)$, one unit of positive charge is equivalent to two units of negative charge (and vice-versa), and hence we could also interpret our illustrations as representing the directed flow of two units of negative charge. This results in a difficulty distinguishing between branching points and monopoles.

\begin{figure}
\includegraphics[width=0.8\linewidth]{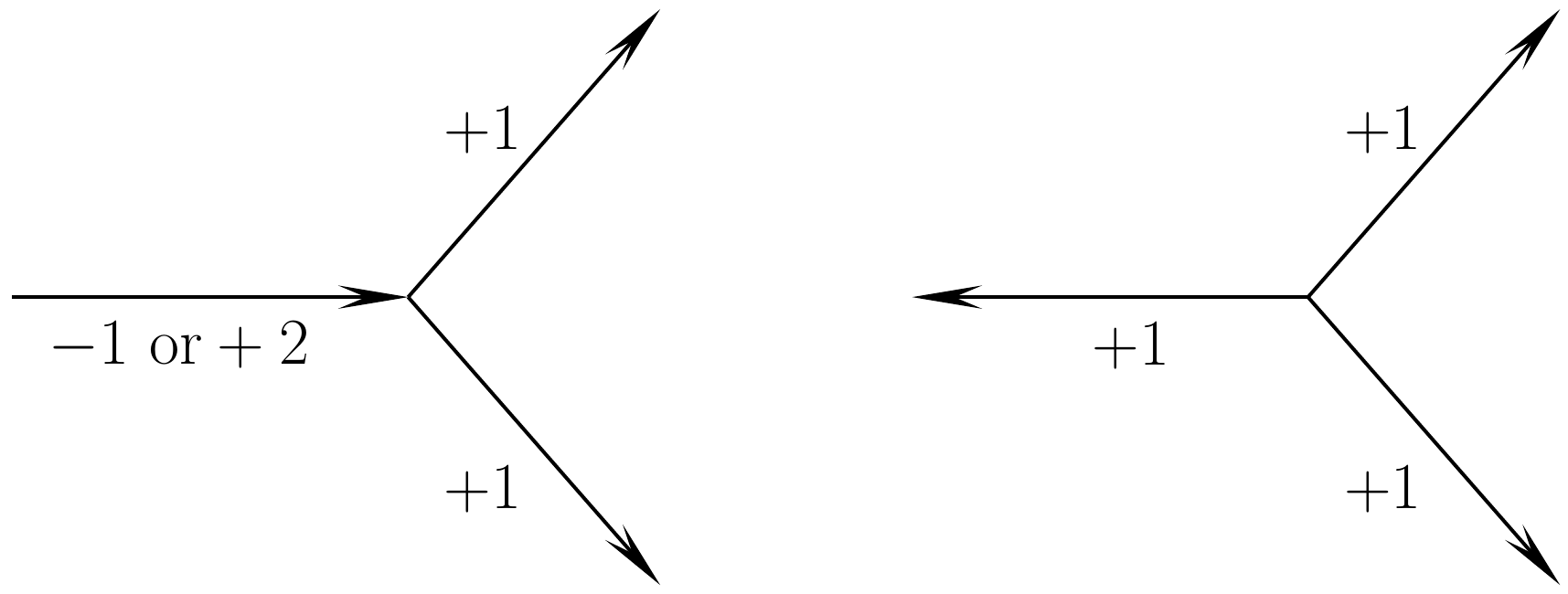}
\caption{\label{fig:VortexBranching} A vortex branching point with centre charge $+2$ flowing into a vertex (left) is equivalent to a vortex monopole with charge $+1$ flowing out of the vertex (right). The arrows indicate the direction of flow for the labelled charge. Our illustrations adopt $m_k\in\lbrace -1,0,+1 \rbrace$ and jets denoting the oriented flow of a positive unit of charge ($m=+1$).}
\end{figure}

This ambiguity is highlighted in Fig.~\ref{fig:VortexBranching}, where we see the equivalence between a branching point and a monopole. For the remainder of this work we will refer to intersections of three or five vortices as branching points rather than monopoles, as the terms are interchangeable without a strict vortex charge limit.
\begin{table}[t]
  \caption{\label{tab:BP}A summary of the possible number of centre vortices piercing a 3D
    cube centred on $\tilde{x}$ and the interpretation of such points.}
  \begin{ruledtabular}
    \begin{tabular}{ c p{7cm} }
      % \hline
      $N_\text{cube}(\tilde{x})$ & Interpretation\\[0.1cm]
      \hline \vspace{-0.2cm}\\
      0 & No vortices present. \\[0.1cm]
      1 & Terminating vortex,
          forbidden by conservation of centre charge. \\[0.1cm]
      2 & Vortex line flowing through the cube. \\[0.1cm]
      3 & Simple three-way branching point. \\[0.1cm]
      4 & Vortex self-intersection. \\[0.1cm]
      5 & Complex five-way branching point. \\[0.1cm]
      6 & Vortex self-intersection or double branching. \\[0.1cm]
    \end{tabular}
  \end{ruledtabular}
\end{table}

Branching points are well defined only on a 3D lattice slice, rather than the full 4D
lattice~\cite{Spengler:2018dxt}. They can be identified at sites $\tilde{x}$ on the dual
lattice by counting the number of vortex lines piercing the elementary 3D cube around
$\tilde{x}$, denoted $N_\text{cube}(\tilde{x})$. $N_\text{cube}(\tilde{x})$ then takes
values from 0 to 6. The interpretation of each value of $N_\text{cube}(\tilde{x})$ is
summarised in Table~\ref{tab:BP}.
% \pagebreak

\begin{figure}
  \includegraphics[width=0.9\linewidth]{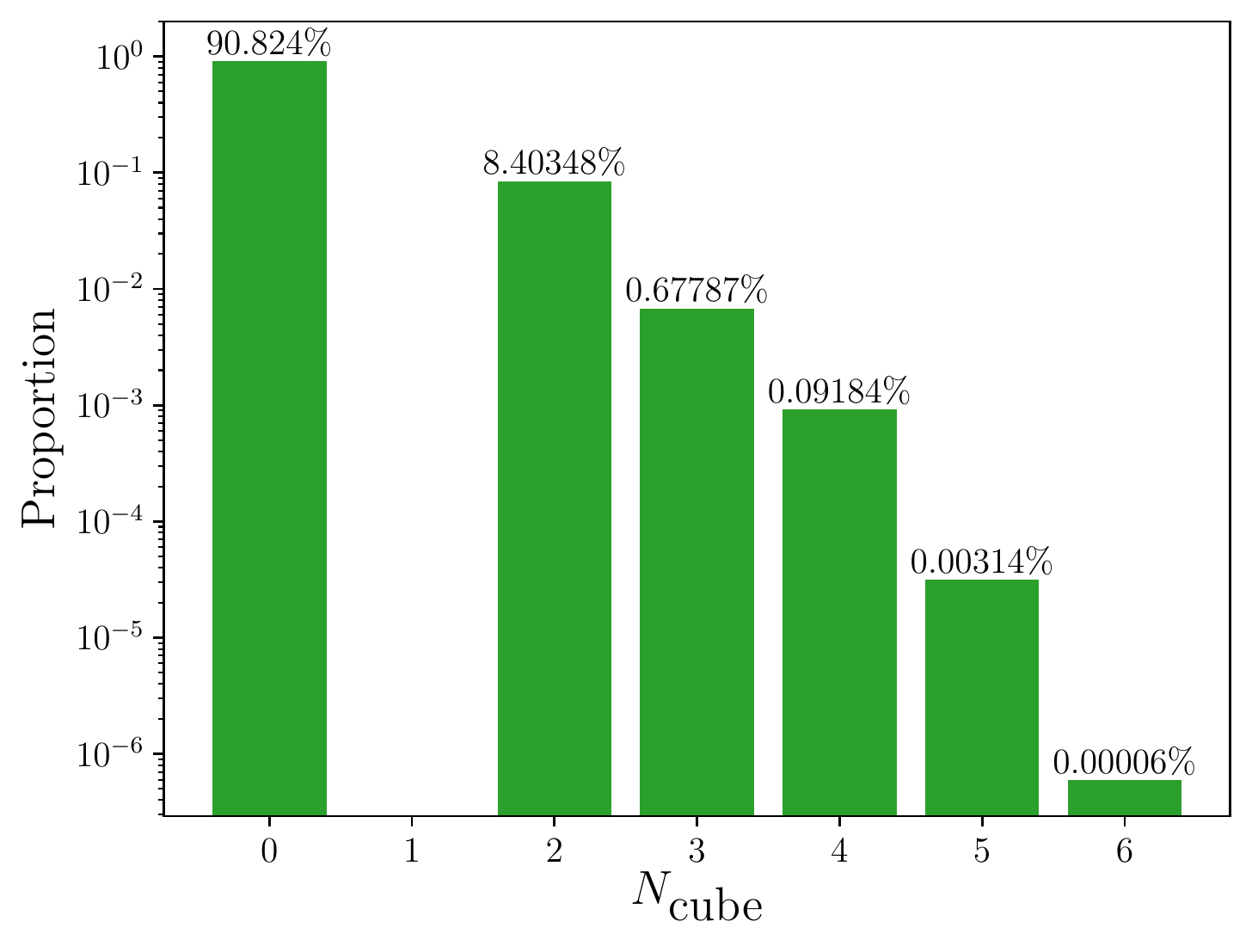}
  \caption{\label{fig:BPHist} The ensemble average of the number of vortices piercing each 3D cube. As it is necessary to preserve continuity of the vortex flux, we see that there are no cubes with one vortex piercing them. The largest vortex contribution is from $N_\text{cube}=2$, arising from vortices propagating without branching or touching. We also see that $N_\text{cube}=3$ branching points dominate the $N_\text{cube}=5$ branching points.}
\end{figure}

The distribution of $N_\text{cube}(\tilde{x})$ over our ensemble is shown in Fig.~\ref{fig:BPHist}. As required, we observe that $N_\text{cube}(\tilde{x}) = 1$ points are not present. Branching points correspond to $N_\text{cube}(\tilde{x}) = 3,~5$, or $6$. As some $N_\text{cube}(\tilde{x}) = 6$ points cannot be unambiguously classified as branching points, they are excluded from our subsequent branching point analysis. This is an acceptable exclusion, as we can see from Fig.~\ref{fig:BPHist} $N_\text{cube}(\tilde{x}) = 6$ points make up only $0.00006\%$ of the total number of 3D cubes in our ensemble, whereas $N_\text{cube}(\tilde{x}) = 3,~5$ branching points are far more prevalent. Thus, we will only consider $N_\text{cube}(\tilde{x}) = 3,~5$ branching points in the following sections.

The fact that branching points are only well defined on 3D slices can be understood by considering the implication of a branching into the fourth dimension. In this case one would observe two vortex jets emerging from or converging into a 3D cube. However, this situation does not occur in our visualisations as is required for conservation of flux lines.

It is clear from our visualisations and the data in Fig.~\ref{fig:BPHist} that branching points occur frequently in the confining phase, with an average of $110(14)$ branching points per 3D slice. This corresponds to a physical density of $\rho_\text{BP} = 3.5(5)\,\text{fm}^{-3}$. Work presented in Refs. ~\cite{Langfeld:2003ev, Spengler:2018dxt} confirms that indeed the branching point density possesses the correct scaling behaviour over different values of $\beta$ governing the lattice spacing such that $\rho_\text{BP}$ may be considered a physical quantity. Further discussion of branching points and their relationship with topological charge is presented in Sec.~\ref{sec:BP}.
\section{Space-Time Oriented Vortices}\label{sec:STOVortices}
For each link in a given 3D slice there are two additional plaquettes that lie in the $x_i - t$ plane, pointing in the positive and negative time directions. Vortices associated with space-time oriented plaquettes contain information about the way the line-like vortices evolve with time, or equivalently, how the vortex surfaces appear in four dimensions.

\begin{figure}
  \includegraphics[width=\linewidth]{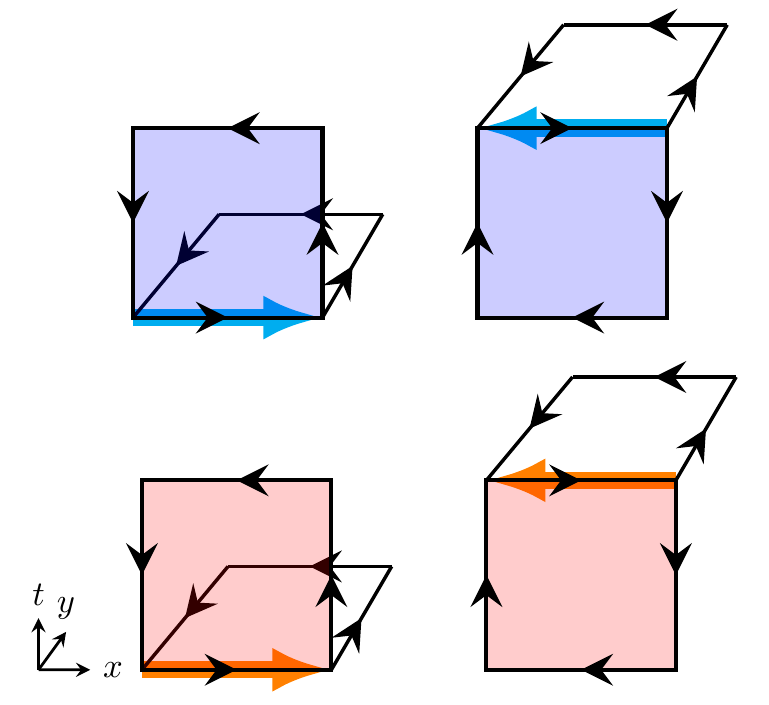}
  \caption{\textit{Top:} A $+1$ vortex in the forward (left)/backward (right) $x-t$ plane
    (shaded blue) will be plotted as a cyan arrow in the $\pm\hat{x}$ direction
    respectively. \textit{Bottom:} A $-1$ vortex in the forward (left)/backward (right)
    $x-t$ plane (shaded red) will be plotted as an orange arrow in the $\pm\hat{x}$
    direction respectively.}
  \label{fig:TimeVortices}
\end{figure}

\begin{figure}
  \includegraphics[width=\linewidth]{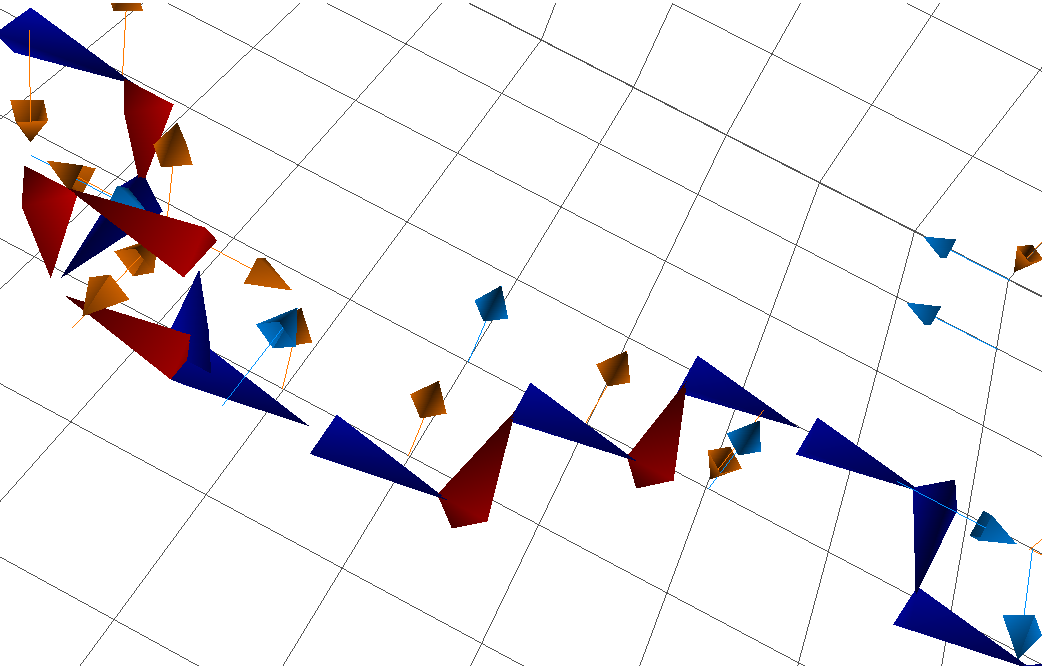}
  \caption{\label{fig:PlaqLinkHighlight}On the $t=1$ time slice, the flow of $m=+1$ centre charge is illustrated by the jets, and the spatial links indicate the presence of centre vortices in the suppressed time direction. These indicator links show how the jets will evolve through the suppressed Euclidean time direction. Rendering conventions are described in the text.}
\end{figure}

In a given 3D slice we only have access to one link associated with a space-time oriented vortex, and as such we plot an arrow along this link to indicate its association with this vortex. Considering the four-dimensional Levi-Cevita tensor, we adopt the following plotting convention for these space-time oriented vortices:
\begin{itemize}[leftmargin=*,itemsep=2pt,labelsep=12pt]
\item[]  {\makebox[3cm]{\stackanchor{$+1$ vortex,}{forward in time,}\hfill} $\implies$ \stackanchor{cyan arrow,}{positively oriented,}}
\item[]  {\makebox[3cm]{\stackanchor{$+1$ vortex,}{backward in time,}\hfill} $\implies$ \stackanchor{cyan arrow,}{negatively oriented,}}
\item[]  {\makebox[3cm]{\stackanchor{$-1$ vortex,}{forward in time,}\hfill}  $\implies$ \stackanchor{orange arrow,}{positively oriented,}}
\item[]  {\makebox[3cm]{\stackanchor{$-1$ vortex,}{backward in time,}\hfill} $\implies$ \stackanchor{orange arrow,}{negatively oriented.}}
\end{itemize}
These conventions are shown diagrammatically in Fig.~\ref{fig:TimeVortices}. Utilising these conventions, the first time slice now contains temporal information as highlighted in Fig.~\ref{fig:PlaqLinkHighlight}. The full 3D models are difficult to interpret in full as a 2D image, however the interactive 3D models for the first two time slices are available in Figs. ~S-3, S-4.
% we see that the first two time slices now appear as Figs.~\ref{fig:PlaqLinkT01}, \ref{fig:PlaqLinkT02}.

%

\begin{figure}
\centering
\subfloat[\label{fig:VortexArrows1}$t=1$]{
\includegraphics[width=0.45\linewidth, height=0.13\textheight]{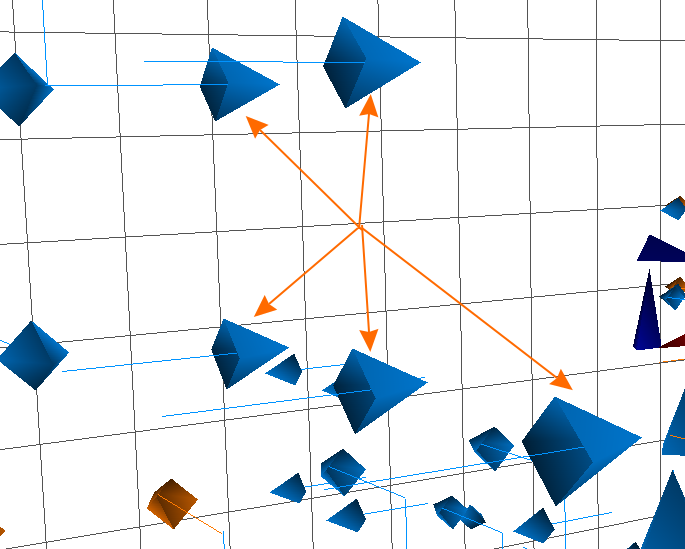}
}
\hfill
\subfloat[\label{fig:VortexArrows2}$t=2$]{
\includegraphics[width=0.45\linewidth, height=0.13\textheight]{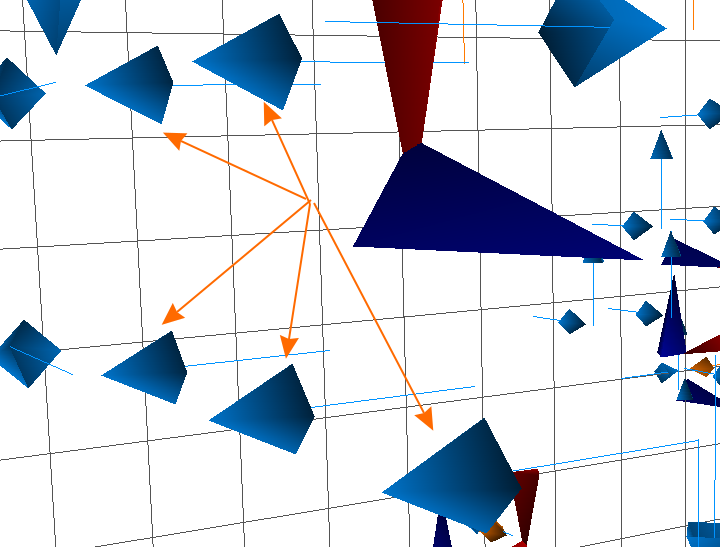}
}
\caption[The change in space-time oriented vortices as we step through time.]{\label{fig:VortexArrows}Space-time oriented vortices changing as we step through time. We observe the space-time indicator links change direction, however the phase (colour) of the vortex remains the same.}
\end{figure}

As we step through time, we expect to see the positively oriented space-time vortex indicator links retain their colour but swap direction as they transition from being forwards in time to backwards in time, as shown in Fig.~\ref{fig:VortexArrows} and in the views ``Forward/backward arrows'' in Figs.~S-3 and S-4.

\begin{figure}
  \includegraphics[width=\linewidth]{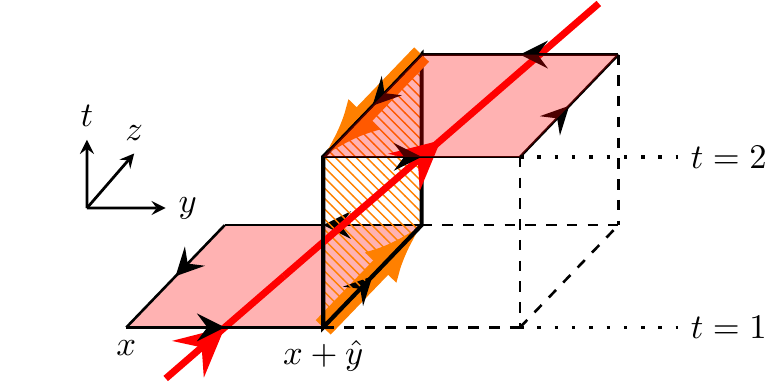}
  \caption{An example of a spatially-oriented vortex at the space-time coordinate $x$ moving one plaquette between time-slices. The solid red line indicates how the flow of vortex charge pierces between time slices. By our visualisation conventions, the shaded red plaquettes would have a spatially-oriented jet plotted in the suppressed $\hat{x}$ direction. Space-time vortices are illustrated by the orange indicator links belonging to the space-time plaquette. We observe that spatially-oriented vortices move in the time direction (hidden in our 3D models), perpendicular to the indicator link.}
  \label{fig:SimpleMotion}
\end{figure}

The space-time oriented indicator links act as predictors of vortex line motion between slices. The simplest case of vortex motion is shown diagrammatically in Fig.~\ref{fig:SimpleMotion}. The shaded red plaquettes indicate the location of a spatially-oriented vortex which would be plotted in the suppressed $\hat{x}$ direction, and the red line demonstrates how the centre charge pierces between the two time slices. This figure demonstrates a spatially-oriented vortex shifting one lattice spacing in the $\hat{y}$ direction between time slices. For a vortex located at $x$ and pointing in the $\pm \hat{x}$ direction, this motion will be indicated by an orange indicator link on the $Z_z(x+\hat{y})$ link. Thus we see that spatially-oriented vortices move in a direction perpendicular to both the jet and the indicator link.

\begin{figure}
\centering
\subfloat[\label{fig:VortexLineMotion1}$t=1$]{
\includegraphics[width=0.45\linewidth, height=0.15\textheight]{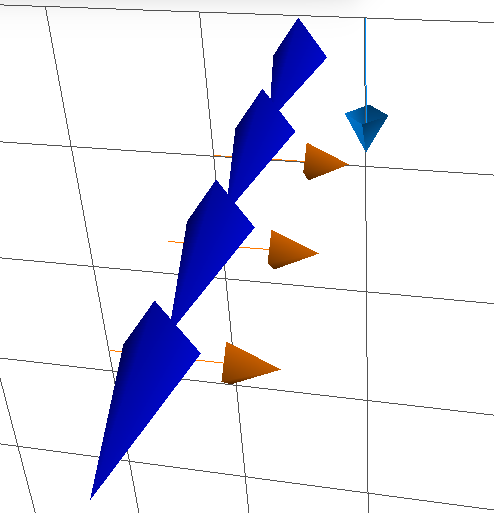}
}
\hfill
\subfloat[\label{fig:VortexLineMotion2}$t=2$]{
\includegraphics[width=0.45\linewidth, height=0.15\textheight]{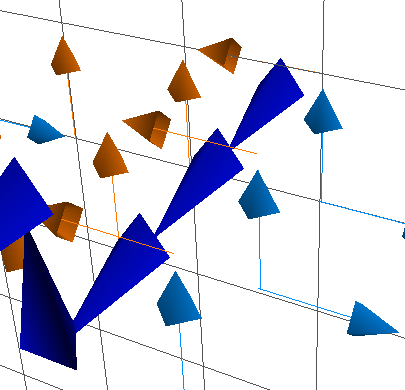}
}
\caption[A second example of space-time oriented vortices predicting the motion of the spatially-oriented vortices.]{\label{fig:VortexLineMotion}An example of space-time oriented vortices predicting the motion of the spatially-oriented vortices. Here we see the $m=+1$ (blue) vortex line transition one lattice spacing down as we step from $t=1$ to $t=2$. Note that the orange space-time vortex indicator links have changed direction.}
\end{figure}

To see this predictive power in action, consider Fig.~\ref{fig:VortexLineMotion}. Here we see in Fig.~\ref{fig:VortexLineMotion1} a line of three $m=+1$ spatially-oriented vortices each with an associated $m=-1$ space-time oriented vortex below them. As we step to $t=2$ in Fig.~\ref{fig:VortexLineMotion2} we observe the space-time oriented arrows change direction, and the spatially-oriented vortex line shifts one lattice spacing down in the direction perpendicular to the indicator links, such that the space-time oriented vortices are now above them.

\begin{figure}
  \centering
  \subfloat[\label{fig:VortexMotion1}$t=1$]{
    \includegraphics[width=0.45\linewidth, height=0.3\textheight]{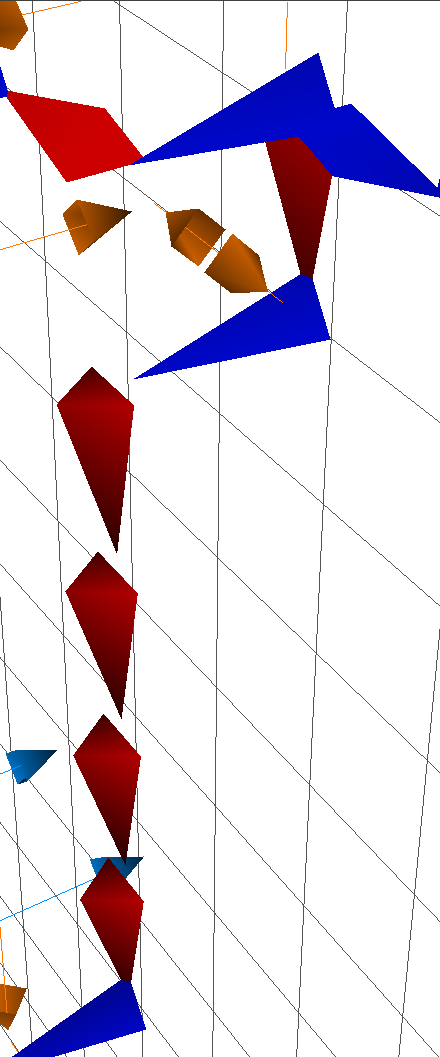}
  }
  \hfill
  \subfloat[\label{fig:VortexMotion2}$t=2$]{
    \includegraphics[width=0.45\linewidth, height=0.3\textheight]{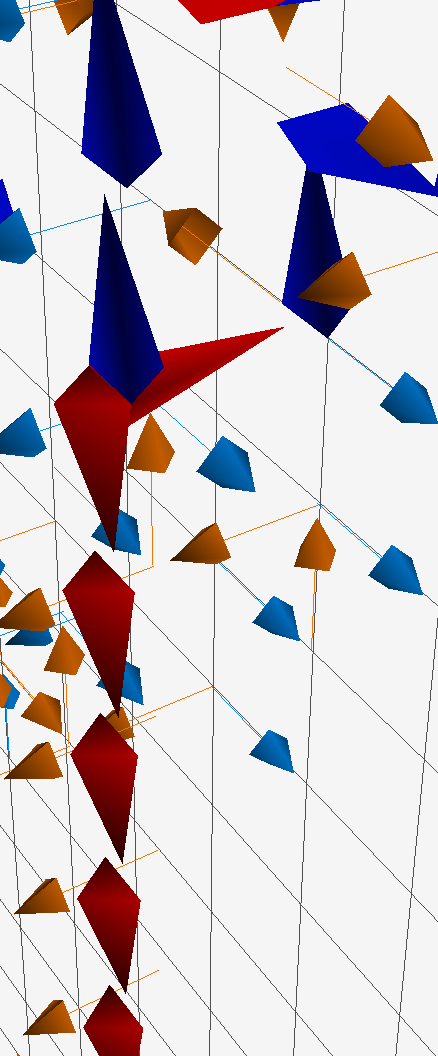}
  }
  \caption[An example of space-time oriented vortices predicting the motion of the
  spatially-oriented vortices.]{\label{fig:VortexMotion}An second example of space-time oriented
    vortices predicting the motion of the spatially-oriented vortices. We observe the $-1$
    (red) vortex line with no associated space-time vortex indicator links remains
    stationary as we transition from $t=1$ to $t=2$. However, the branching point with
    associated space-time vortex indicator link moves down and to the left during the
    transition.}
\end{figure}

Another example of space-time oriented vortices predicting the motion of spatially-oriented vortex lines is shown in Fig.~\ref{fig:VortexMotion}. In Fig.~\ref{fig:VortexMotion1}, we observe a line of four $m=-1$ (red) spatially-oriented vortices with no space-time oriented links associated with them, indicating that this line should remain fixed as we step through time. Alternatively, towards the top of the red line we observe a branching point with two associated $-1$ space-time indicator arrows. The forward-oriented arrow indicates that this branching point will move. That is, the sheet piercing the $t=1$ slice is generating nontrivial space-time vortices as it proceeds to pierce the $t=2$ slice. Observing the same region at $t=2$ in Fig.~\ref{fig:VortexMotion2}, we see that this is precisely what occurs. The vortex line has remained fixed, whereas the branching point has shifted. This vortex motion can also be examined in the views ``Vortex line behaviour'' in Figs.~S-3 and S-4.

\begin{figure}
\centering
\subfloat[\label{fig:VortexLadder1}$t=1$]{
\includegraphics[width=0.45\linewidth]{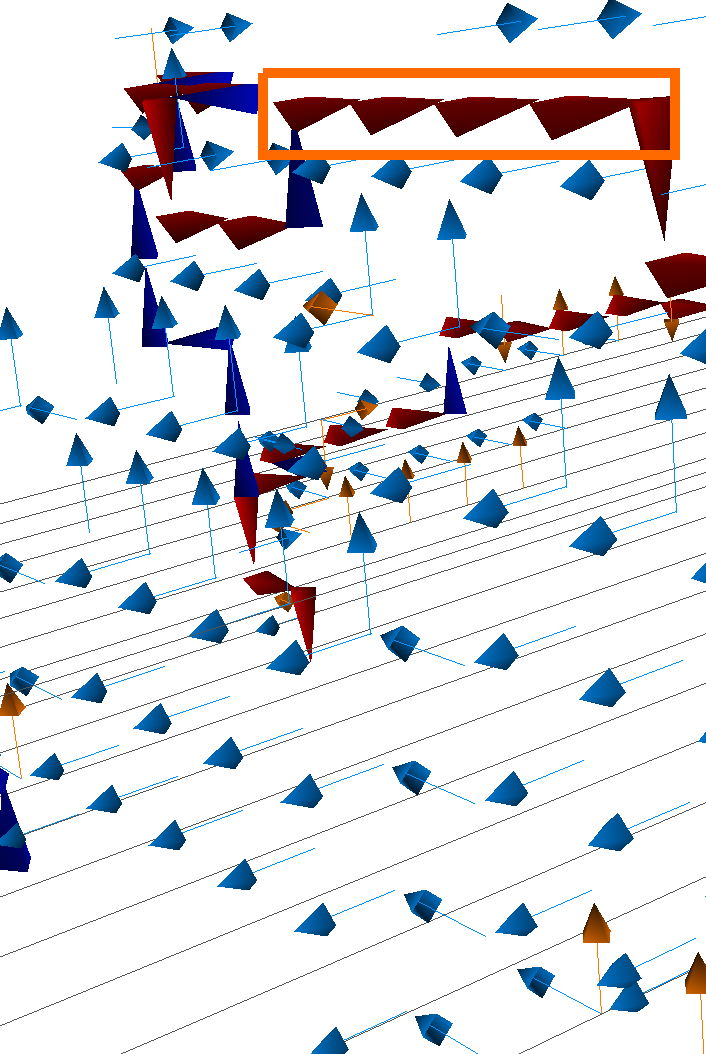}
}
\hfill
\subfloat[\label{fig:VortexLadder2}$t=2$]{
\includegraphics[width=0.45\linewidth]{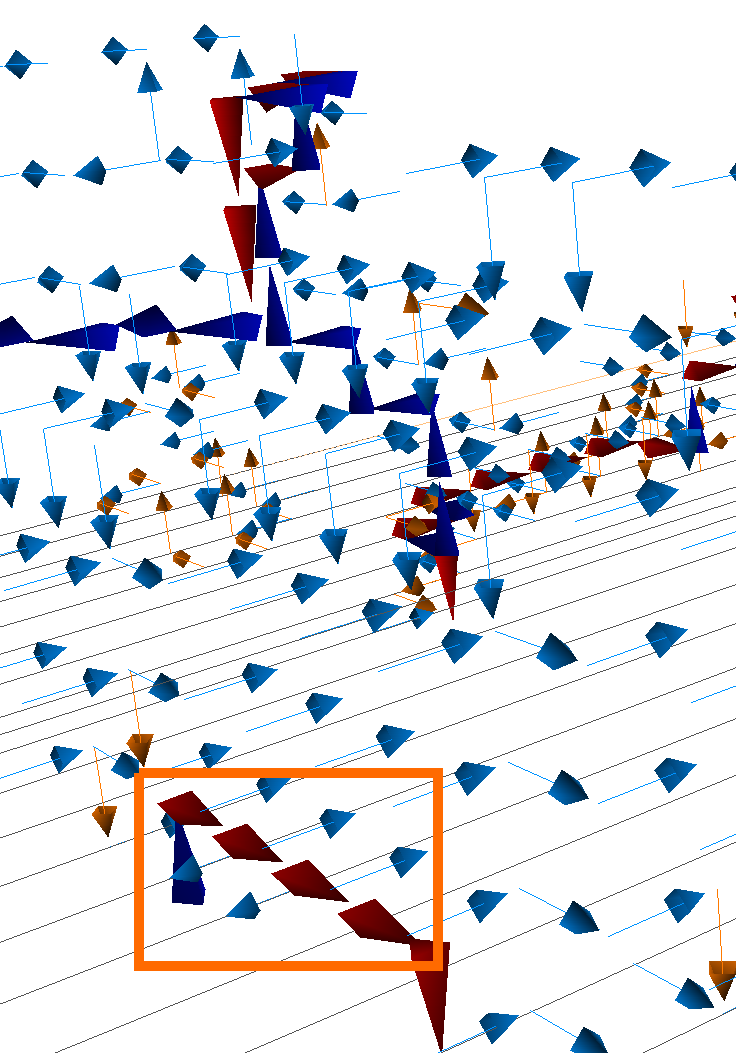}
}
\caption{\label{fig:VortexLadder}An example of a sheet of space-time oriented vortices predicting the motion of spatially-oriented vortices over multiple lattice sites from $t=1$ to $t=2$. The highlighted line of red vortices flows along the sheet of cyan time-oriented indicator links.}
\end{figure}

The cases presented in Fig.~\ref{fig:VortexLineMotion} and Fig.~\ref{fig:VortexMotion} are ideal, where the spatially-oriented vortex lines shift only one lattice spacing between time slices. However, it is frequently the case where the spatially-oriented vortices shift multiple lattice spacings per time step, as demonstrated in Fig.~\ref{fig:VortexLadder}. In Fig.~\ref{fig:VortexLadder1}, we observe a large sheet of space-time oriented vortices with a line of spatially oriented vortices above them. As we transition to $t=2$ in Fig.~\ref{fig:VortexLadder2}, the line is carried along the sheet and now appears at the bottom.

\begin{figure}
  \centering
  \includegraphics[width=\linewidth]{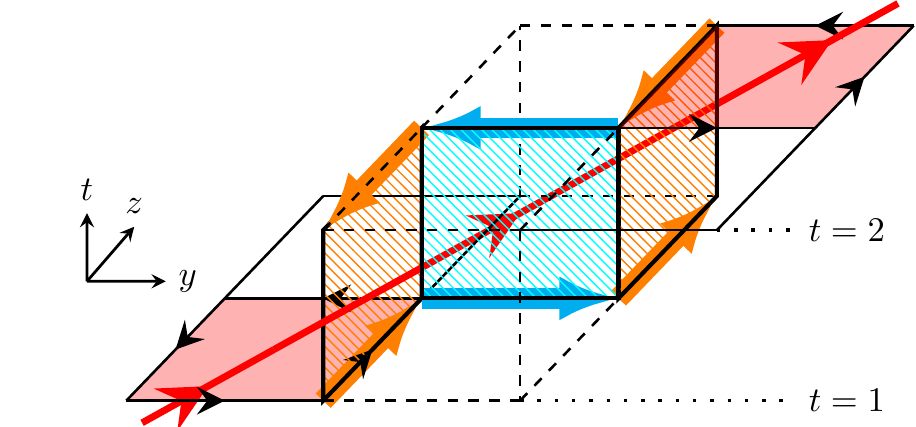}
\caption{\label{fig:ComplexStructure}A demonstration of how spatially-oriented vortices can transition multiple lattice spacings in a single time step. Conventions are the same as in Fig.~\ref{fig:SimpleMotion}}
\end{figure}

To see how this occurs diagrammatically, consider Fig.~\ref{fig:ComplexStructure}. The conventions in this figure are the same as in Fig.~\ref{fig:SimpleMotion}. Within each slice we would observe the space-time oriented links shown; however the spatially-oriented vortex appears to move three plaquettes in one time step. These multiple transitions make it difficult to track the motion of vortices between time slices. However, the space-time oriented vortices remain a useful tool for understanding how centre vortices evolve with time. Note that if a spatially-oriented vortex has no associated space-time oriented vortices then it is guaranteed to remain stationary. In this respect, the lack of space-time oriented vortices is also a valuable indicator of vortex behaviour.


\section{Topological Charge}\label{sec:TopChargeVis}
We now wish to explore the relationship between vortices and topological charge. The topological charge density is given by
\begin{equation}\label{eq:TopQ}
q ( x ) = \frac { 1 } { 32 \pi ^ { 2 } } \, \epsilon ^ { \mu \nu \rho \sigma } \, \Tr\left( F _ { \mu \nu }( x ) \, F _ { \rho \sigma }(x) \right) \, .
\end{equation}
Topological charge is calculated on the lattice by evaluation of clover terms $C_{\mu\nu}$. The simplest $1\times 1$ clover term is given by~\cite{BilsonThompson:2002jk}
\begin{align}\label{eq:Clover}
C_{\mu\nu}(x) =& \frac{1}{4} \operatorname{Im}\left[P_{\mu\nu}(x) +  P_{\mu\nu}(x-\hat{\mu})\right.\nonumber\\
&\left. + P_{\mu\nu}(x - \hat{\nu}) + P_{\mu\nu}(x - \hat{\mu} - \hat{\nu})\right]\, .
\end{align}
From these terms, we obtain
\begin{equation}\label{eq:topq}
q(x) = \frac{1}{32\pi^2}\, \epsilon^{ \mu \nu \rho \sigma } \, \Tr\left( C _ { \mu \nu }( x ) \, C _ { \rho \sigma }(x) \right) \, .
\end{equation}
To improve the topological charge density calculation by removing higher-order error terms, it is possible to make use of an improved lattice field-strength tensor by taking into account larger Wilson loops in the definition of the clover terms, as described in Ref.~\cite{BilsonThompson:2002jk}. In this paper we make use of the simple $1\times 1$ topological charge definition when analysing projected configurations, as the gauge link information is highly localised around the projected vortex locations. However, for the original and smoothed configurations we instead employ a 5-loop improved definition as it produces more accurate results and shows better convergence to integer values on smoothed configurations~\cite{BilsonThompson:2002jk}.

We calculate the topological charge density on an original lattice configuration after eight sweeps of three-loop $\mathcal{O}(a^4)$-improved cooling~\cite{BilsonThompson:2002jk}. This smoothing is necessary to remove short-range fluctuations and associated large perturbative renormalisations, but is a sufficiently low number of sweeps so as to minimally perturb the configuration. Topological charge density obtained after minimal over-improved stout-link smearing is explored in Sec.~\ref{sec:Correlation}.

\begin{figure}
  \centering
\includegraphics[width=\linewidth]{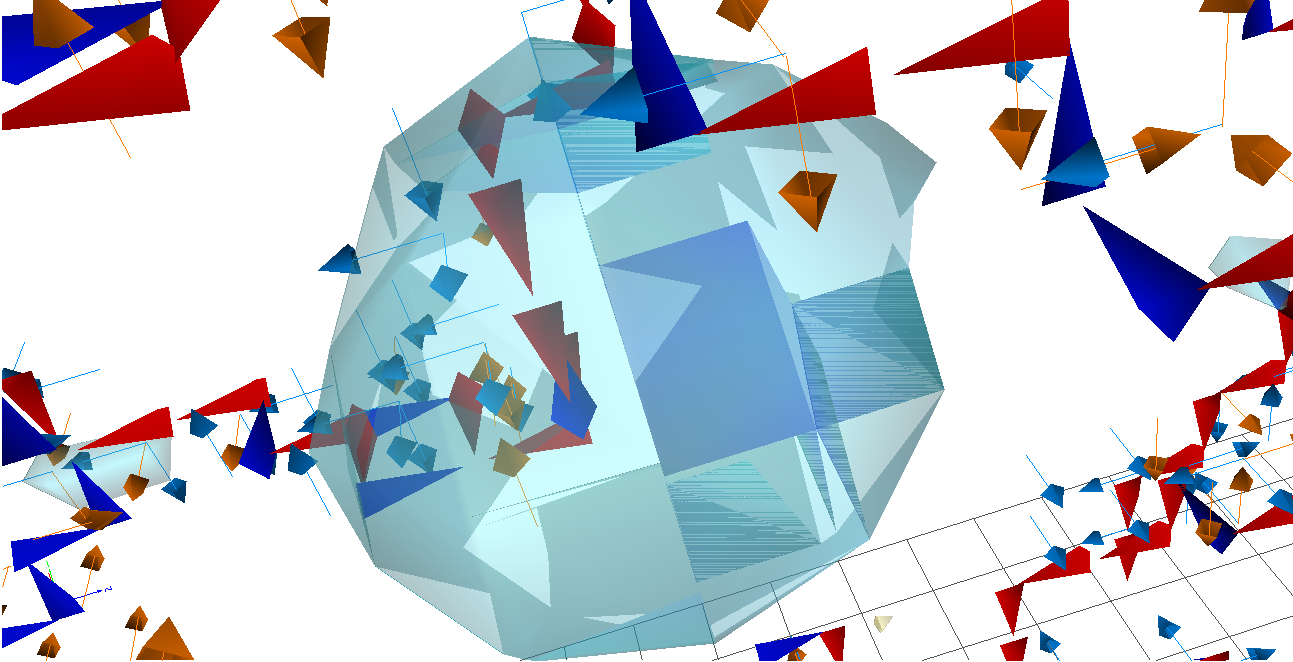}

\caption{\label{fig:PlaqLinkQT01Highlight}Regions of high topological charge density are rendered
  as translucent blue ($q(x)<0$) and yellow ($q(x)>0$) volumes, overlaying the $t=1$
  slice.}

  % \vspace*{\floatsep} \includegraphics[width=\linewidth]{PlaqLinkTopQ_CFG95_T02.png}

  % \caption{\label{fig:PlaqLinkQT02}Regions of high topological charge density overlaying
  %   the $t=2$ slice. Conventions are as in Fig.~\ref{fig:PlaqLinkQT01}
  %   (\textbf{Interactive})}
\end{figure}

We plot regions of positive topological charge density in yellow, and regions of negative topological charge density in blue, with a colour gradient to indicate the magnitude. Only topological charge density of sufficient magnitude is plotted to better emphasise regions of significance. Overlaying the topological charge density visualisation onto our previous 3D models, we obtain the visualisation shown in Fig.~\ref{fig:PlaqLinkQT01Highlight}. The full 3D models are available in Figs.~S-5 and S-6.

\begin{figure}
  \includegraphics[width=\linewidth]{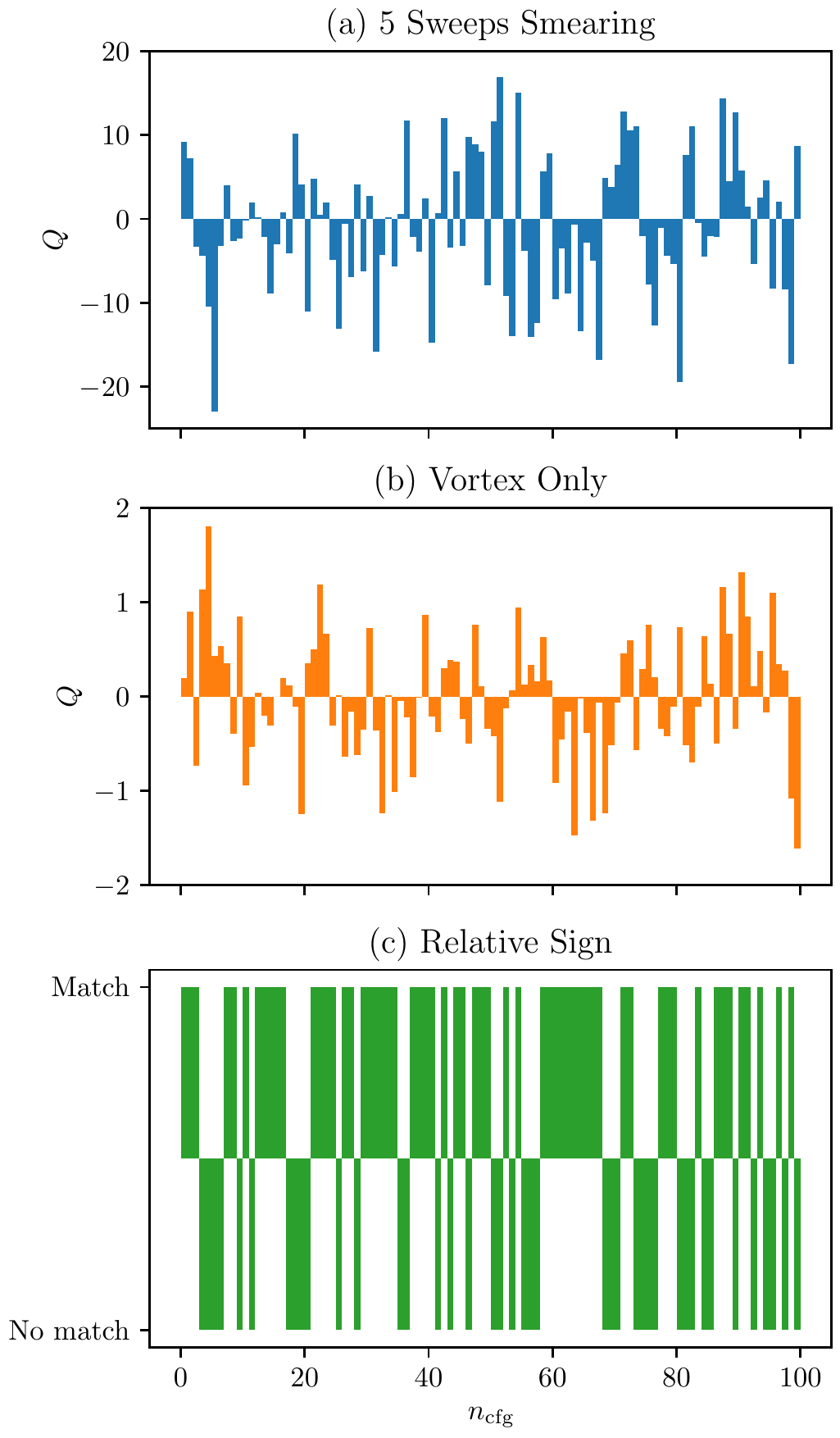}
  \caption{A histogram of total topological charge $Q$ (a) using the gluonic
      definition after 5 sweeps of over-improved stout-link smearing and (b) direct centre
      projection from the original gauge fields. It is apparent that singular points
      following centre projection do not preserve the total topological charge (note
      also the scale change between (a) and (b)). Panel (c) shows whether the sign matches
      between plots (a) and (b); again it is apparent that there is little correlation.}
  \label{fig:TopQHist}
\end{figure}

Under centre projection the topological charge changes notably, as might be expected for a local operator. Fig.~\ref{fig:TopQHist} shows a histogram of the total topological charge $Q$ across the ensemble obtained with the gluonic definition after five sweeps of over-improved stout-link smearing. This is compared to $Q$ obtained from the singular points of the projected configurations. Clearly, the topological charge is not preserved. We also check in panel (c) of Fig. ~\ref{fig:TopQHist} whether the relative sign of the topological charge calculated on each configuration is the same. Here we also observe little correlation in the relative sign.

Observing the percolation of nontrivial centre vortices in the context of topological
charge density provides new insight into the instability of instanton-like objects to
centre-vortex removal~\cite{Trewartha:2015ida}. We can quantitatively evaluate the
correlation between the locations of centre vortices and the regions of significant
topological charge density obtained from the vortex-only configurations by using the
measure

\begin{equation}
C = V\frac{\sum_x |q(x)|\,L(x)}{\sum_x |q(x)|\,\sum_x L(x)} - 1\, ,
\label{eq:Correlation}
\end{equation}
where $V$ is the lattice volume, and
\begin{equation}
L(x) =
\begin{cases}
1\, , & \text{\stackanchor{Vortex associated with any}{plaquette touching $x$,}}\\[10pt]
0\, , & \text{otherwise,}\,
\end{cases}
\end{equation}
contains information from the full 4D volume. This method of constructing $L(x)$ allows for a single vortex to result in multiple non-zero $L(x)$ locations. However, $L(x)$ is defined in this way so that vortex information associated with plaquettes is shifted to the regular lattice, allowing it to be compared with the topological charge density. A value of $C=0$ indicates no correlation. $C<0$ and $C>0$ indicate anti-correlation or correlation respectively.

We can also compare the results of this calculation to the maximally correlated value for $C$, which can be obtained by postulating that all $x$ for which $L(x) = 1$ correlate to the $\sum_x L(x)$ highest values of $|q(x)|$, denoted $|q_i|$. As we are assuming perfect correlation, $L(x) = 1$ for all $i$, and hence the numerator of Eq.~\ref{eq:Correlation} reduces to a sum over $|q_i|$. Hence,
\begin{equation}
  \label{eq:CIdeal}
  C_\text{Ideal} = V \frac{\sum_{i=1}^{N} |q_i|}{\sum_x |q(x)|\,\sum_x L(x)} - 1\, ,
\end{equation}
where $N$ is the number of sites with $L(x) = 1$. By evaluating $C/C_\text{Ideal}$ for each configuration, we obtain a normalised measure ranging between $0$ and $1$ for positively correlated quantities. Averaging over our configurations, we can make use of $\overline{C/C_\text{Ideal}}$ to quantitatively express the correlation strength between $L(x)$ and $|q(x)|$.

Evaluating $C/C_\text{Ideal}$ and averaging over our ensemble of 100 configurations provides $\overline{C/C_\text{Ideal}}=0.672(6)$. Thus, there is a significant correlation between the positions of vortices and topological charge density. The small uncertainty also indicates that this correlation is consistent across the ensemble.

% This result indicates a low degree of correlation, most likely arising from the the high density of centre vortices. It is common for centre vortices to pass through  regions of significant topological charge density and their removal will spoil the important long-range spacetime-colour field correlations that otherwise protect instanton-like objects under local smoothing algorithms.

\begin{figure}
\includegraphics[width=\linewidth]{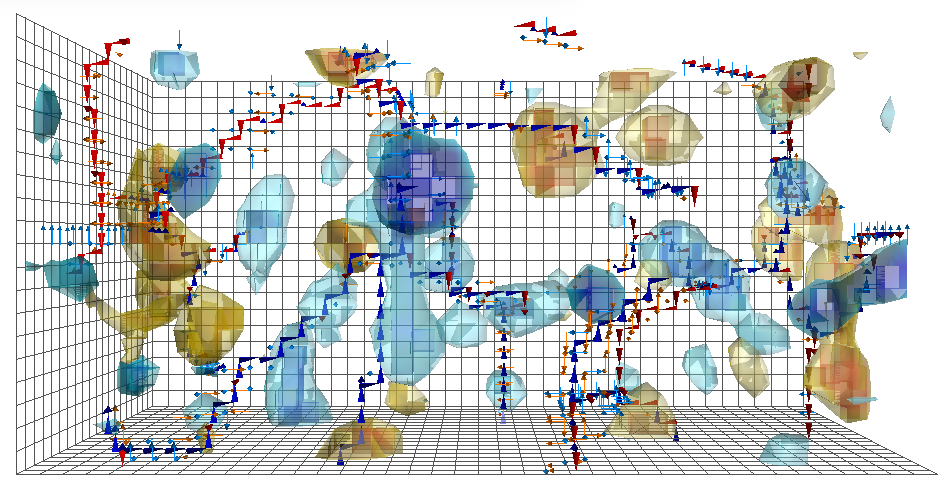}
  \caption{\label{fig:PlaqLinkTopQ_SW8}The centre vortex structure and topological charge density after eight sweeps of cooling, for $t=1$. (\textbf{Interactive})}
\end{figure}

Finally, we visualise the vortex configurations after smoothing to investigate how the vortex structure changes. The results, presented in Fig.~\ref{fig:PlaqLinkTopQ_SW8}, follow eight sweeps of $\mathcal{O}(a^4)$-improved cooling. We note that an enormous amount of the vortex matter is removed. However, it is well established that, under smoothing, the vortex-only configurations retain many of the salient long-range features of QCD~\cite{Trewartha:2017ive, Biddle:2018dtc, Trewartha:2015ida}, suggesting that the removed vortices are in some way irrelevant to these long-range properties.

\section{Singular Points}\label{sec:SP}
Given the presence of the antisymmetric tensor in the definition of topological charge density presented in Eq.~\eqref{eq:TopQ}, it is clear that for there to be nontrivial topological charge present on the projected vortex configurations, we require the vortex field strength to span all four dimensions. This condition is met at \textit{singular points}, where the tangent vectors of the vortex surface span all four dimensions. The contribution to the topological charge from these singular points is discussed in detail in Refs.~\cite{Bruckmann:2003yd,Engelhardt:2010ft,Engelhardt:2000wc,Engelhardt:1999xw}.

\begin{figure}
  \includegraphics[width=0.5\linewidth]{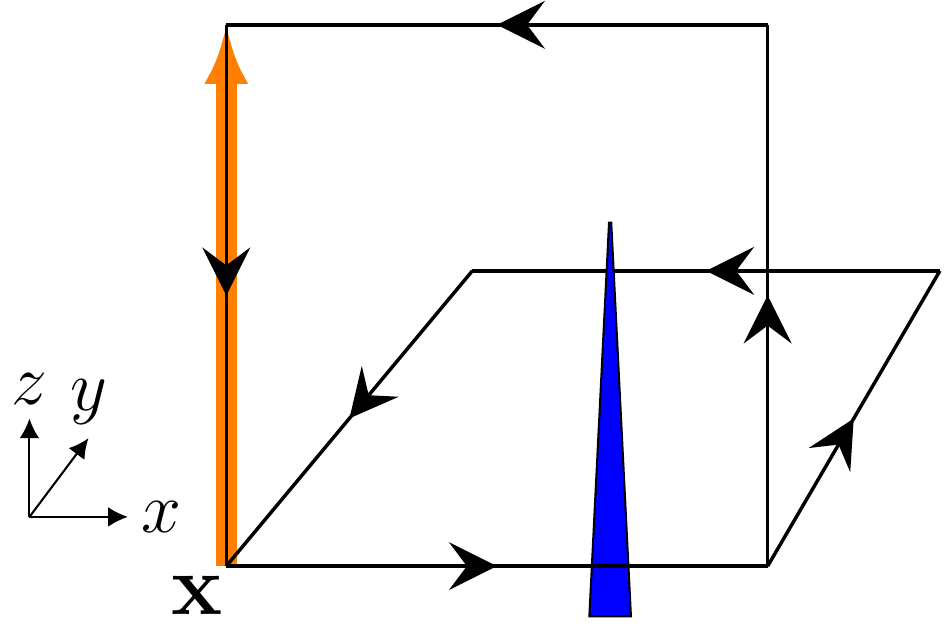}
\caption{\label{fig:SingularPoint} The signature of a singular point, in which the tangent vectors of the vortex surface span all four dimensions. In this case, the blue jet is associated with field strength in the $x-y$ plane, and the orange space-time vortex indicator link is associated with a vortex generating field strength in the $z-t$ plane. Hence, the vortex surface spans all four dimensions.}
\end{figure}

\begin{figure}
\includegraphics[width=\linewidth]{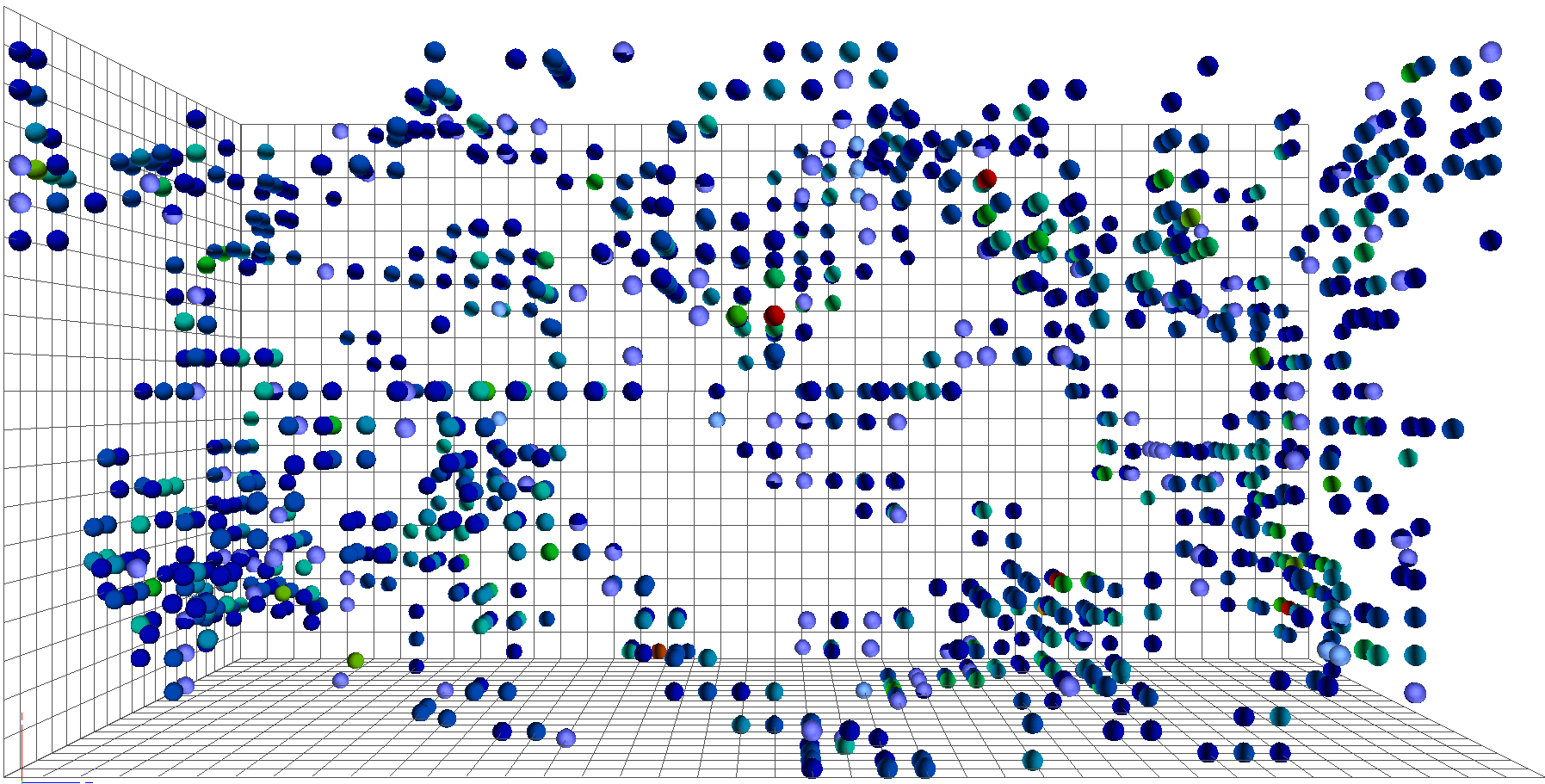}
  \caption{All points on the $t=1$ time slice in which a singular point occurs, i.e. a spatially-oriented vortex jet runs parallel to a space-time oriented vortex indicator link as shown in Fig.~\ref{fig:SingularPoint}. The colour indicates the multiplicity $M$ observed on this slice, with the lowest value in blue ($M=1$) and the highest in red ($M=12$). (\textbf{Interactive})}
  \label{fig:AllSP}
\end{figure}

\begin{figure}
\includegraphics[width=0.7\linewidth]{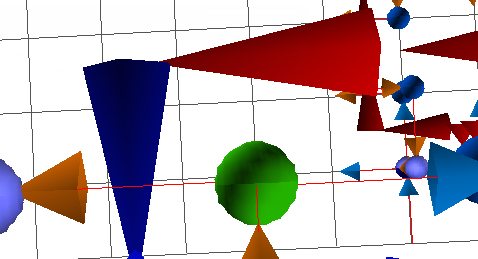}
\caption{\label{fig:SingularPointVis}A singular point (green sphere) resembling the
  structure of Fig.~\ref{fig:SingularPoint}. This singular point is generated by the red
  jet and the orange indicator link running in parallel.}
\end{figure}

In our visualisations, singular points appear as a spatially-oriented vortex running parallel to the link identifier of a space-time oriented vortex, as shown in Fig.~\ref{fig:SingularPoint}. Points satisfying this condition, whilst being difficult to locate by eye in our visualisations of space-time oriented vortices, actually occur frequently, as illustrated in Fig.~\ref{fig:AllSP} (the interactive model is available in Fig.~S-8). At these points we have vortices generating field strength in all four space-time dimensions. An example of a singular point from the visualisation in Fig.~S-4 is shown in Fig.~\ref{fig:SingularPointVis}.

The vortex configuration in Fig.~\ref{fig:SingularPoint} spans all four dimensions because the jet indicates a vortex in the $x-y$ plane generating non-zero field strength $F_{xy}(x)$ and the $z$-oriented indicator link denotes a vortex in the $z-t$ plane, giving rise to non-zero $F_{zt}(x)$. Hence, at the point $x$ the topological charge density can be nontrivial.

Around the lattice site $\mathbf{x}$ in Fig.~\ref{fig:SingularPoint} there are four $x-y$ and four $z-t$ plaquettes, allowing for a multiplicity of 16. As there are three unique combinations of orthogonal planes in 4D ($x-y$ and $z-t$, $x-z$ and $y-t$, $y-z$ and $x-t$), this gives a total maximum multiplicity of 48 for each singular point. However, this maximum is highly unlikely, and the highest multiplicity in the configuration shown in Fig.~\ref{fig:AllSP} is 12. This point is shown in the view ``Maximum multiplicity'' singular point' in Figs.~S-3 and S-8.

\begin{figure}
\includegraphics[width=\linewidth]{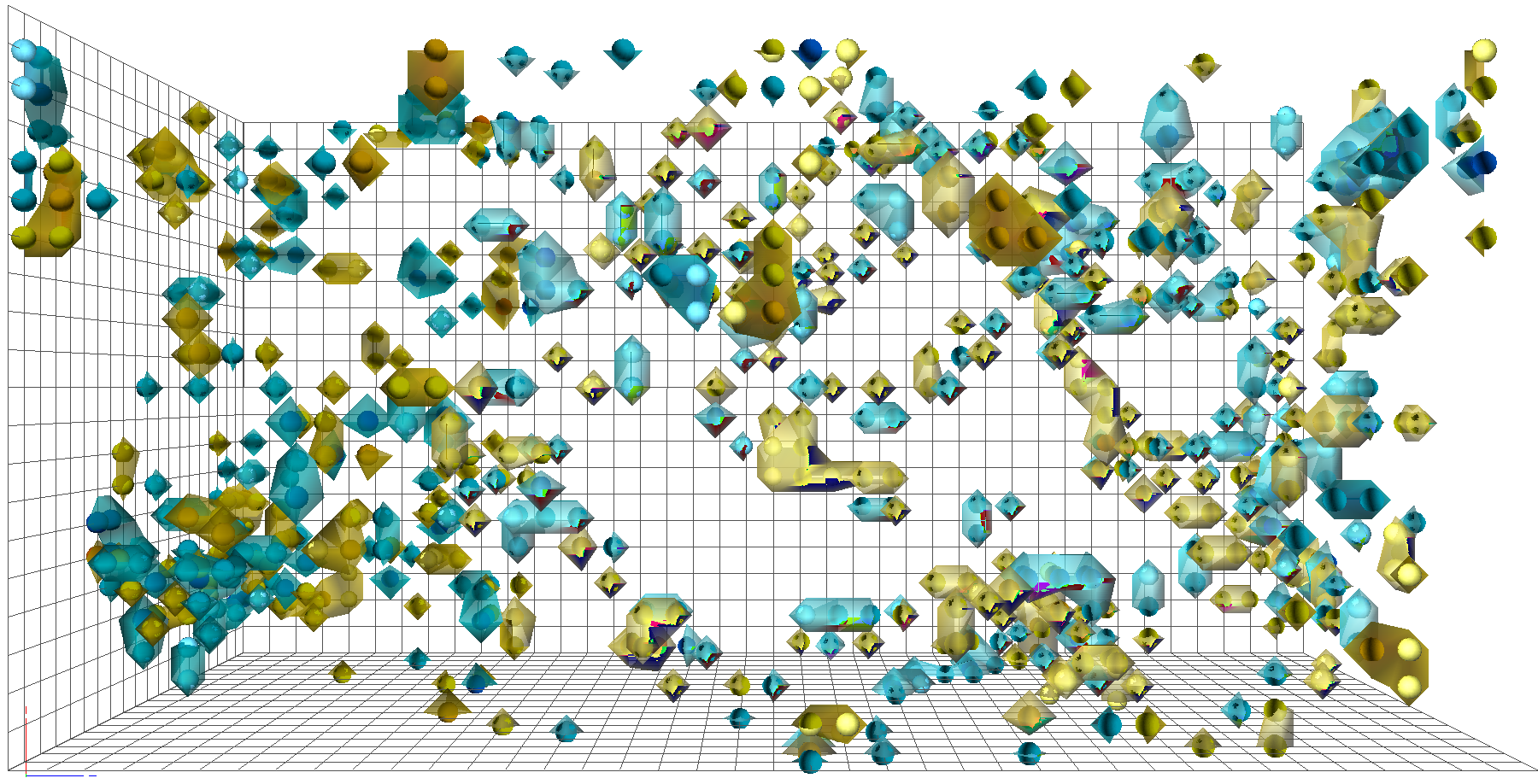}
  \caption{\label{fig:TopQSing}Topological charge density from singular points (shown as
    dots) is compared with topological charge calculated from vortex-only configurations for
    $t=1$. (\textbf{Interactive})}
\end{figure}

We can verify the relationship between singular points and topological charge by identifying vortices satisfying the parallel condition shown in Fig.~\ref{fig:SingularPoint} and plotting these points against the results of the topological charge calculation performed on the projected vortex-only configurations. As seen in Fig.~\ref{fig:TopQSing}, when we apply the correct sign to the odd index permutations we observe that there is perfect agreement between the location of singular points and the identified topological charge.

To quantify the correlation between $q(x)$ and singular points, we make use of a measure similar to that defined in Eq.~\eqref{eq:Correlation},
\begin{equation}
C = V\frac{\sum_x |q(x)|\,L_s(x)}{\sum_x |q(x)|\,\sum_x L_s(x)} - 1\, .
\end{equation}
However, we redefine our identifier $L(x)$ to be
\begin{equation}
L_s(x) =
\begin{cases}
1\, , & \text{singular point at $x$,}\\[10pt]
0\, , & \text{otherwise.}\,
\end{cases}
\end{equation}
In the case of singular points and $|q(x)|$ obtained from the projected configurations, we expect that the obtained correlation will be identical to the ideal value, calculated in the same manner as Eq.~\ref{eq:CIdeal}. This is indeed what we observe, with $\overline{C/C_\text{Ideal}} = 1$. In Sec.~\ref{sec:Correlation} we will make use of this measure again to examine the correlation between singular points and different topological charge density calculated prior to centre vortex projection where the expected values are less apparent.

\section{Branching points}\label{sec:BP}
\begin{figure}
\includegraphics[width=\linewidth]{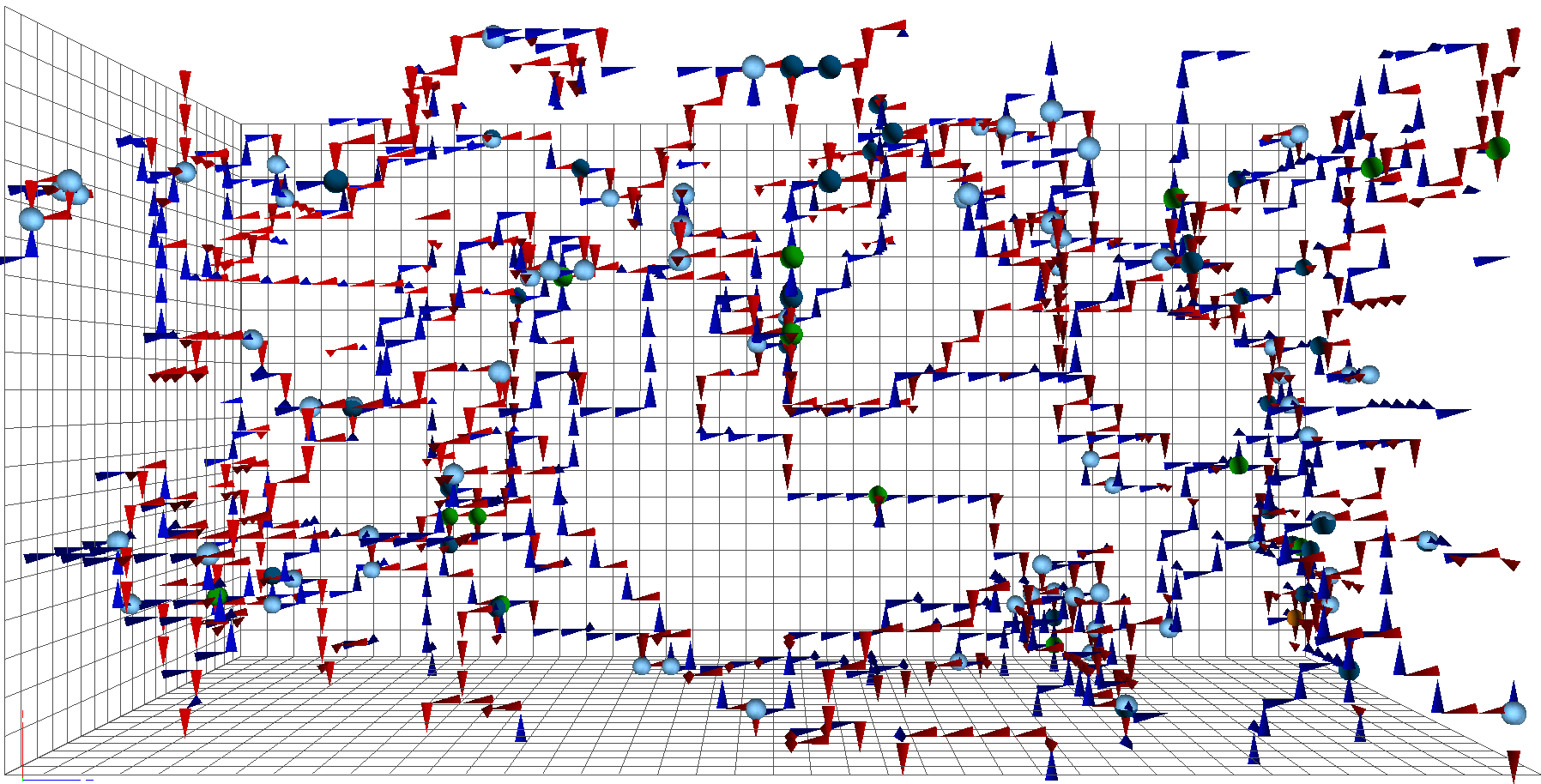}
  \caption{\label{fig:BPVO}Points with two or more vortices piercing a 3D cube are shown
    on the $t=1$ time slice. The number of vortices piercing a cube is denoted by the
    colour: blue = 3, green = 4, orange = 5, red = 6. Whilst there are no red points
    present in this slice, they occur rarely on other slices. (\textbf{Interactive})}
\end{figure}

As mentioned in Sec.~\ref{sec:SOVortices}, the $SU(3)$ gauge group permits branching points, identified as the intersection of three or five spatially-oriented vortices in an elementary 3D cube. The branching points are highlighted for $t=1$ on our sample configuration in Fig.~\ref{fig:BPVO}.

\begin{figure}
  \includegraphics[width=0.8\linewidth]{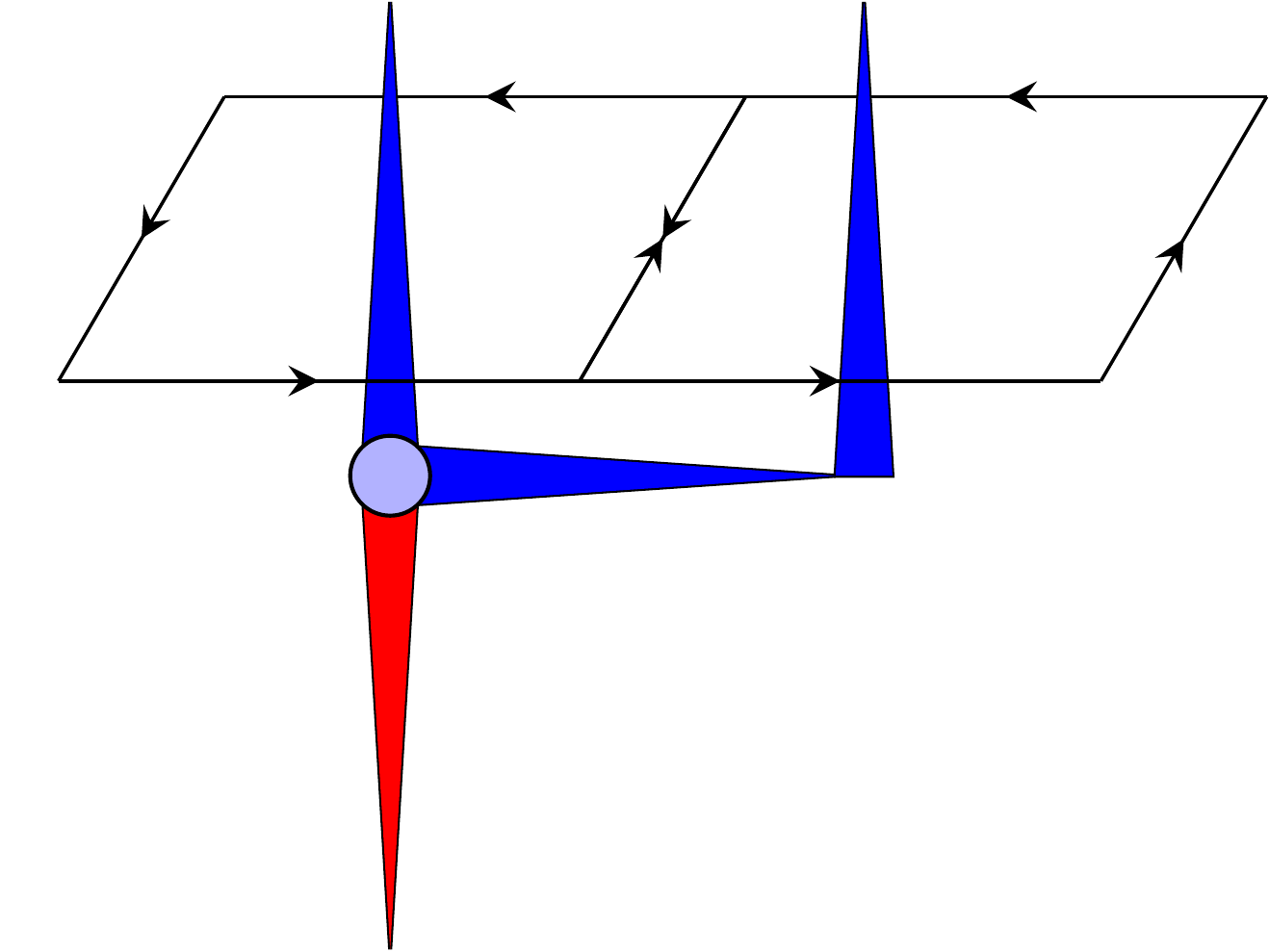}
  \caption{An example of vortex branching generating a region of high topological charge
    by piercing two out of the four plaquettes that make up a clover.}
  \label{fig:BPExample}
\end{figure}

Branching points are of particular interest as they are important for generating regions of high topological charge density on the projected vortex configurations. To understand the reason why, consider a clover term $C_{\mu\nu}$ as defined in Eq.~\eqref{eq:Clover}. On a projected configuration, each of the four imaginary parts of the plaquettes in Eq.~\eqref{eq:Clover} can take one of three possible values: $\pm \sqrt{3}/2$ or $0$. Topological charge density of the lowest magnitude will be given by each of the orthogonal clover terms in Eq.~\eqref{eq:topq} contributing $\pm \sqrt{3}/2$, either because the remaining plaquettes in each clover do not contribute, or because they contribute but cancel due to opposing signs. To obtain larger values of $|q(x)|$, it is therefore necessary for multiple plaquettes in at least one of the clover terms to contribute both nontrivially and with the same sign so that the magnitude of the topological charge density increases above the lowest nontrivial value. This is equivalent to requiring that multiple vortex jets pierce the clover parallel to each other, such that they form a pattern like that shown in Fig.~\ref{fig:BPExample}. To conserve the vortex flux, the  configuration in Fig.~\ref{fig:BPExample} is most simply achieved by placing a branching point immediately below the two parallel vortices. Hence, there is reason to suspect that branching points may be well-correlated with regions of high topological charge.

\begin{figure}
\includegraphics[width=\linewidth]{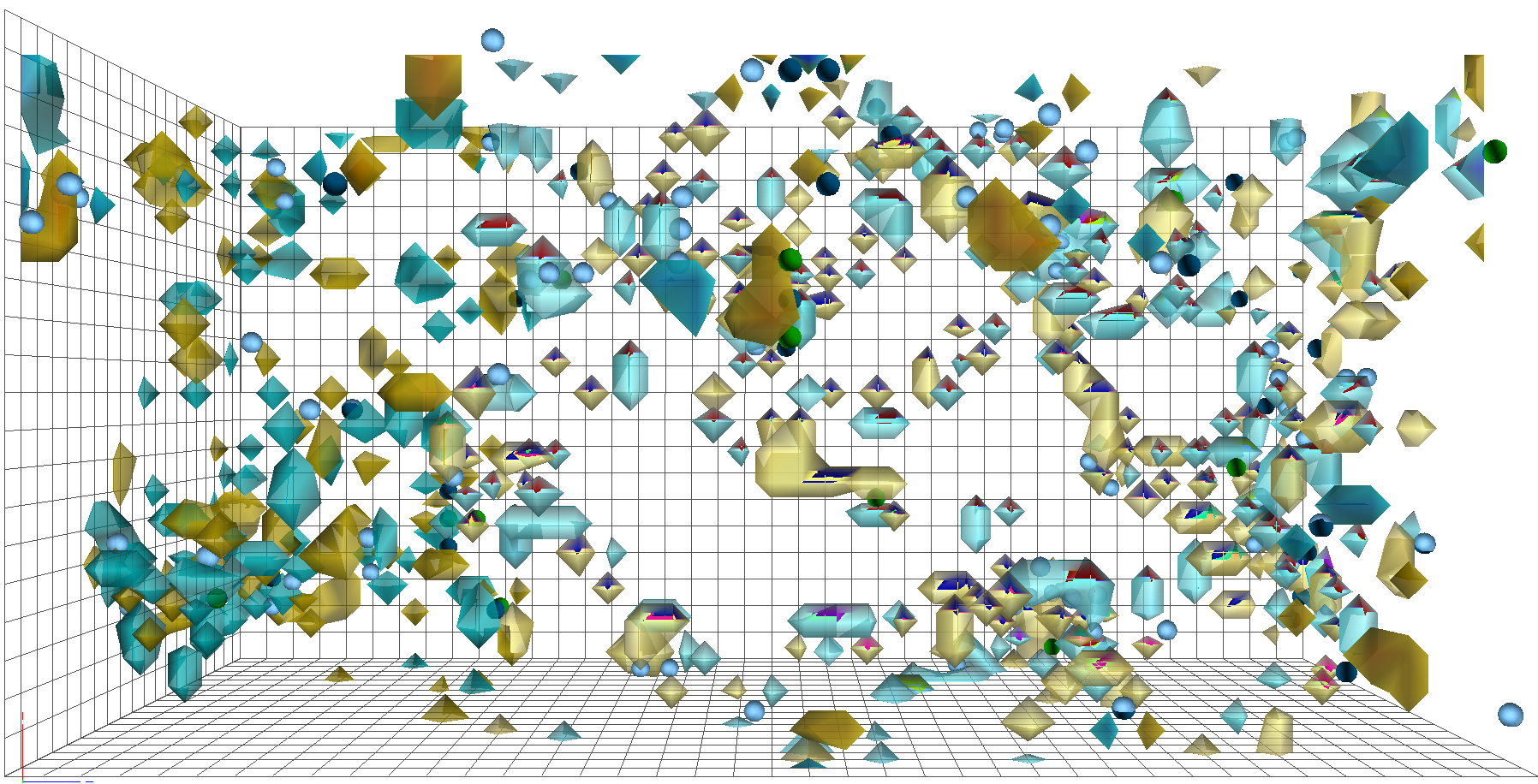}
  \caption{\label{fig:BPq} Branching points (dots) plotted alongside the topological
    charge density from the projected vortex configurations. It can be observed that the
    branching points are almost always neighbouring topological charge
    density. (\textbf{Interactive})}
\end{figure}

\begin{figure}
  \includegraphics[width=0.5\linewidth]{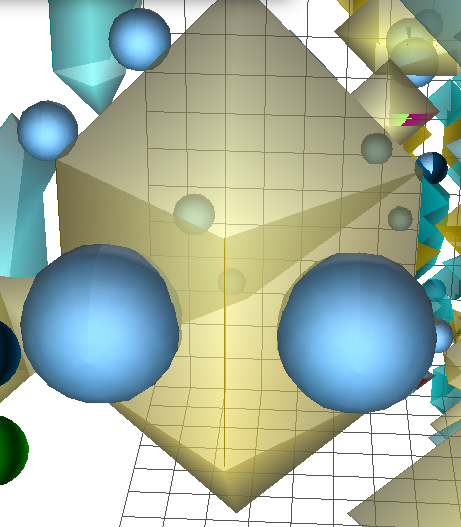}
  \caption{\label{fig:BP&Q}An example of two branching points and their associated topological charge density.}
\end{figure}

The argument made above by no means claims that branching points \textit{must} be associated with large values of $|q(x)|$, as there are most certainly alternative vortex arrangements that will lead to the same values. For example, a branching point could generate two parallel vortex lines that then continue parallel to one another for some distance, generating topological charge density away from the original branching point. Or alternatively, two separate vortex lines could come close to one another, running parallel without the need for any local branching point. Thus, the correlation between large values of $|q(x)|$ and branching points is not expected to be perfect, however the presence of a correlation provides information on the role of branching points in generating large topological charge density. Inspection of the 3D model in Fig.~\ref{fig:BPq} suggests a significant correlation, as is highlighted in Fig.~\ref{fig:BP&Q} and in the view ``Two branching points and their associated topological charge'' in Fig.~S-11.

To evaluate the correlation numerically we again make use of the measure defined in Eq.~\eqref{eq:Correlation}. As branching points are defined as the intersection of three or five vortices, they exist on the dual 3D lattice of each time-slice. The dual lattice sites are denoted by $\tilde{x}$. For the four unique combinations of three dimensions, $xyz$, $xyt$, $xzt$ and $yzt$, we define our branching point indicator measure as
\begin{equation}
  L_\mu(\tilde{x}) =
  \begin{cases}
    1\, , & \text{\stackanchor{Branching point associated with $\tilde{x}$}{in 3D slices of constant $\mu$,}}\\[10pt]
    0\, , & \text{otherwise}\, .
  \end{cases}
  \label{eq:BPmu}
\end{equation}
The $\mu$ index in Eq.~\eqref{eq:BPmu} indicates which dimension is playing the role of time, i.e. which dimension is not included in the 3D cubes. Similarly, we define $q_\mu(\tilde{x})$ to be the average of the topological charge over each 3D cube around $\tilde{x}$. We then have four correlation measures for each 3D combination that can be averaged over, giving a total correlation of
\begin{equation}
  C = \frac{1}{4} \sum_\mu V\frac{\sum_{\tilde{x}} |q_\mu(\tilde{x})|\,L_\mu(\tilde{x})}{\sum_{\tilde{x}} |q_\mu(\tilde{x})|\,\sum_{\tilde{x}} L_\mu(\tilde{x})} - 1\, .
  \label{eq:Cmu}
\end{equation}
By constructing the ideal correlation as defined in Eq.~\eqref{eq:CIdeal} for each choice of 3D coordinates and averaging as done in Eq.~\eqref{eq:Cmu}, we can also calculate the ideal correlation with which we can compare against.

With this measure now suitably defined, we find that we obtain an ensemble average of $\overline{C/C_\text{Ideal}} = 0.518(7)$. This result indicates that there is a notable correlation between branching points and topological charge density and, as expected, they are not the only source of large topological charge. This result is interesting as it speaks to the tendency of vortex lines to either re-combine or diverge away from branching points, rather than remain in close proximity to one another, which provides an interesting consideration for the construction of $SU(3)$ vortex models such as those presented in Refs.~\cite{Engelhardt:2003wm, Engelhardt:2010ft}.

% \clearpage
\section{Correlation with Topological Charge Density}\label{sec:Correlation}
When considering correlations between vortex matter and topological charge density, it is natural to wonder whether the vortex structures identified on the projected vortex-only configurations correlate to the topological charge density identified on the original configurations. As is well established, to accurately identify topological charge density directly from the lattice gauge links it is necessary to first perform smoothing to filter short-range fluctuations~\cite{Moran:2008ra,BilsonThompson:2003zi}.

To this end, we perform five sweeps of over-improved stout-link smearing, with smearing parameters $\epsilon = -0.25$ and $\rho = 0.06$, to minimally smooth the configurations before extracting the topological charge density~\cite{Moran:2008ra}. To ensure the smoothed configurations maintain information captured in the vortex projection, we also produce smeared configurations that are preconditioned in maximal centre gauge.

\begin{figure}
  \includegraphics[width=\linewidth]{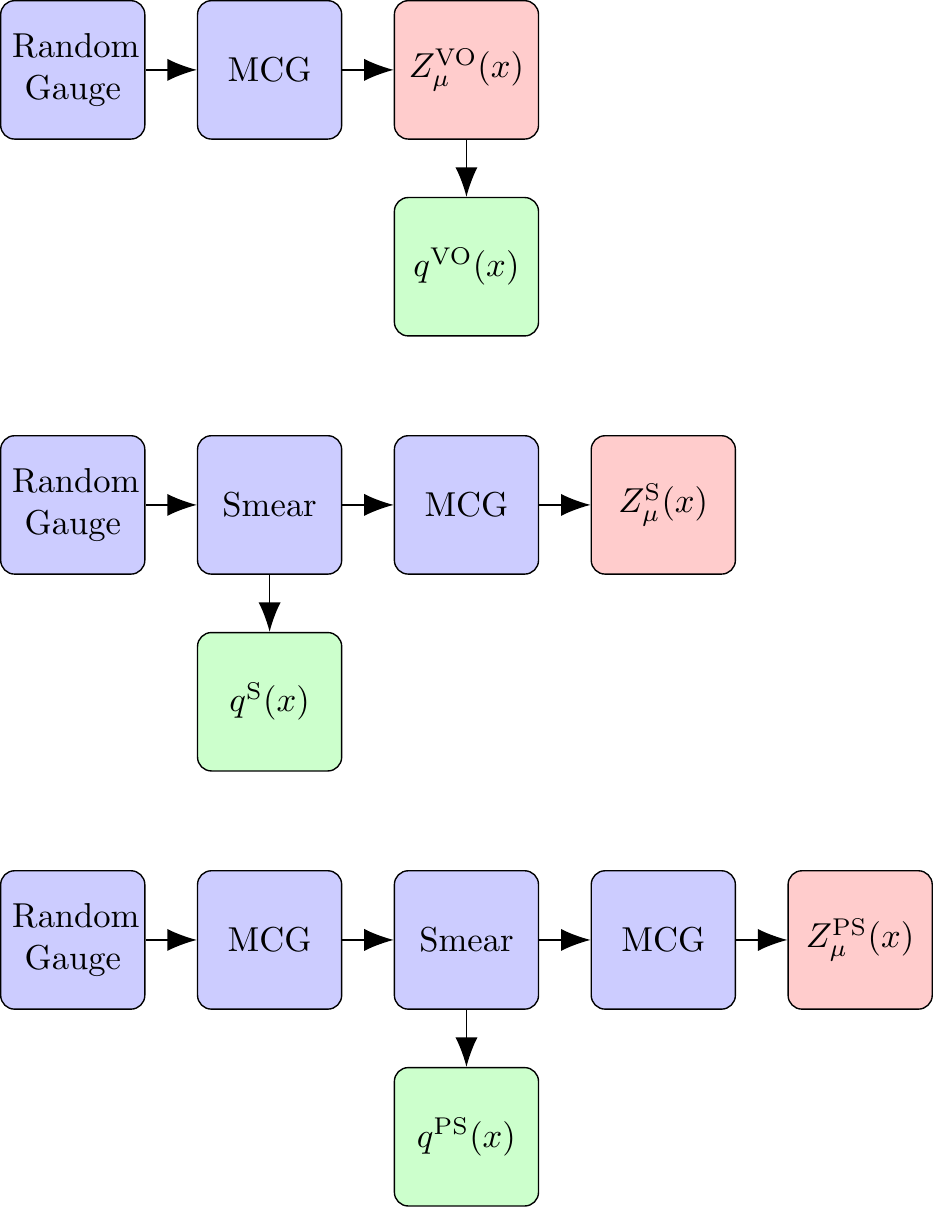}
  \caption{Summary of the processes used to obtain vortex and topological charge density
    configurations. ``MCG'' denotes gauge fixing to maximal centre gauge and ``Smear'' denotes
    application of five sweeps of over-improved stout-link smearing as described in the
    text. From these methods we obtain the vortex only (VO), smeared (S) and
    preconditioned smeared (PS) topological charge and vortex configurations. As the
    topological charge density is gauge invariant, it could equivalently be calculated
    following gauge fixing to maximal centre gauge.}
  \label{fig:FlowChart}
\end{figure}

We also obtain vortices from these smoothed configurations by fixing them to maximal centre gauge and then centre projecting, giving us in total three vortex configurations and three topological charge configurations. The methods by which these ensembles are obtained are summarised in Fig.~\ref{fig:FlowChart}.

\begin{figure*}
  \includegraphics[width=\linewidth]{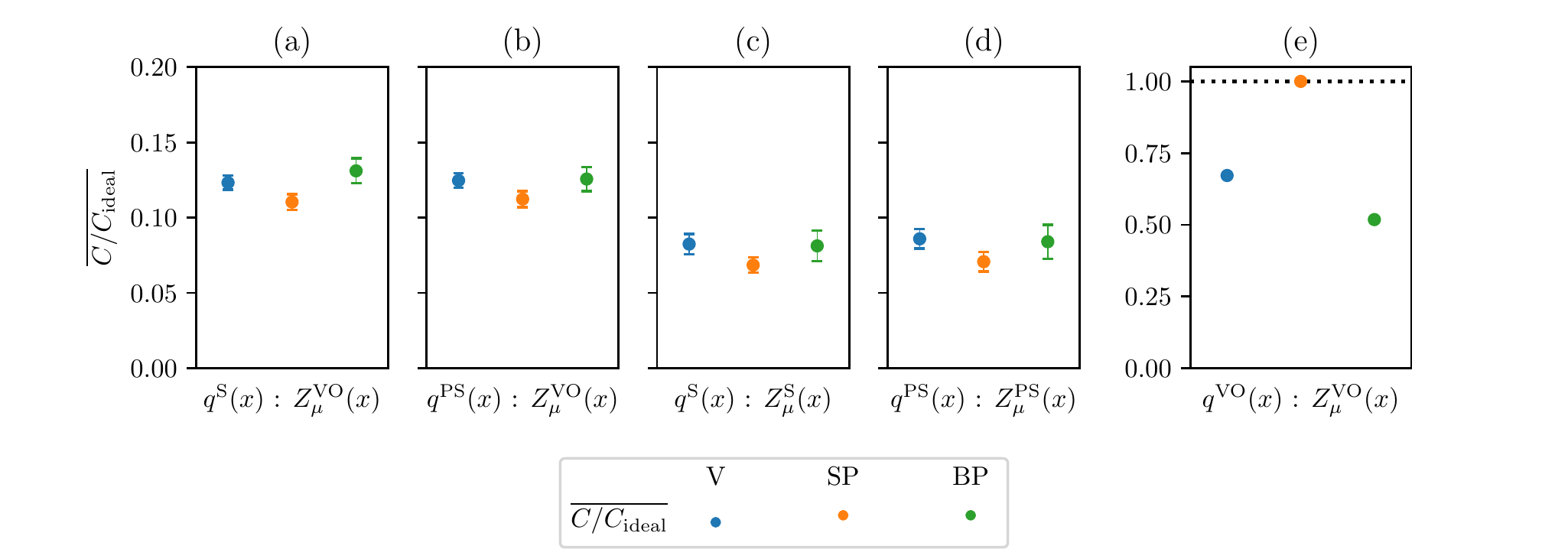}
  \caption{Correlation values for vortices (V), singular points (SP) and branching
      points (BP) obtained from centre projected configurations, $Z_\mu(x)$, with
      topological charge density, $q(x)$, obtained via various means described in the text
      and shown diagrammatically in Fig.~\ref{fig:FlowChart}. Data points indicate
      results for the normalised correlation values, $\overline{C/C_\text{Ideal}}$.}
  \label{fig:AllCorrelation}
\end{figure*}

We now repeat our correlation calculations for the singular points, branching points and the vortices themselves for four new combinations of vortex and topological charge density configurations. These results, as well as the correlation results from the previous sections, are summarised in Fig.~\ref{fig:AllCorrelation}. We see that for all of the new correlations presented, there is a soft correlation between the vortex structures and the topological charge density. Of all the correlations of $q^\text{S}(x)$ or $q^\text{PS}(x)$ with vortex information, the strongest correlation is with the original $Z_\mu^\text{VO}(x)$. It is notable that the branching point correlation is similar to the vortex correlation in panels (a) and (b) of Fig.~\ref{fig:AllCorrelation}. These configurations best represent the physical gauge fields with minimal smoothing to extract the topological charge, and as such this correlation has notable implications for the significance of branching points in regards to generating regions of significant topological charge density.

\begin{figure}
\includegraphics[width=\linewidth]{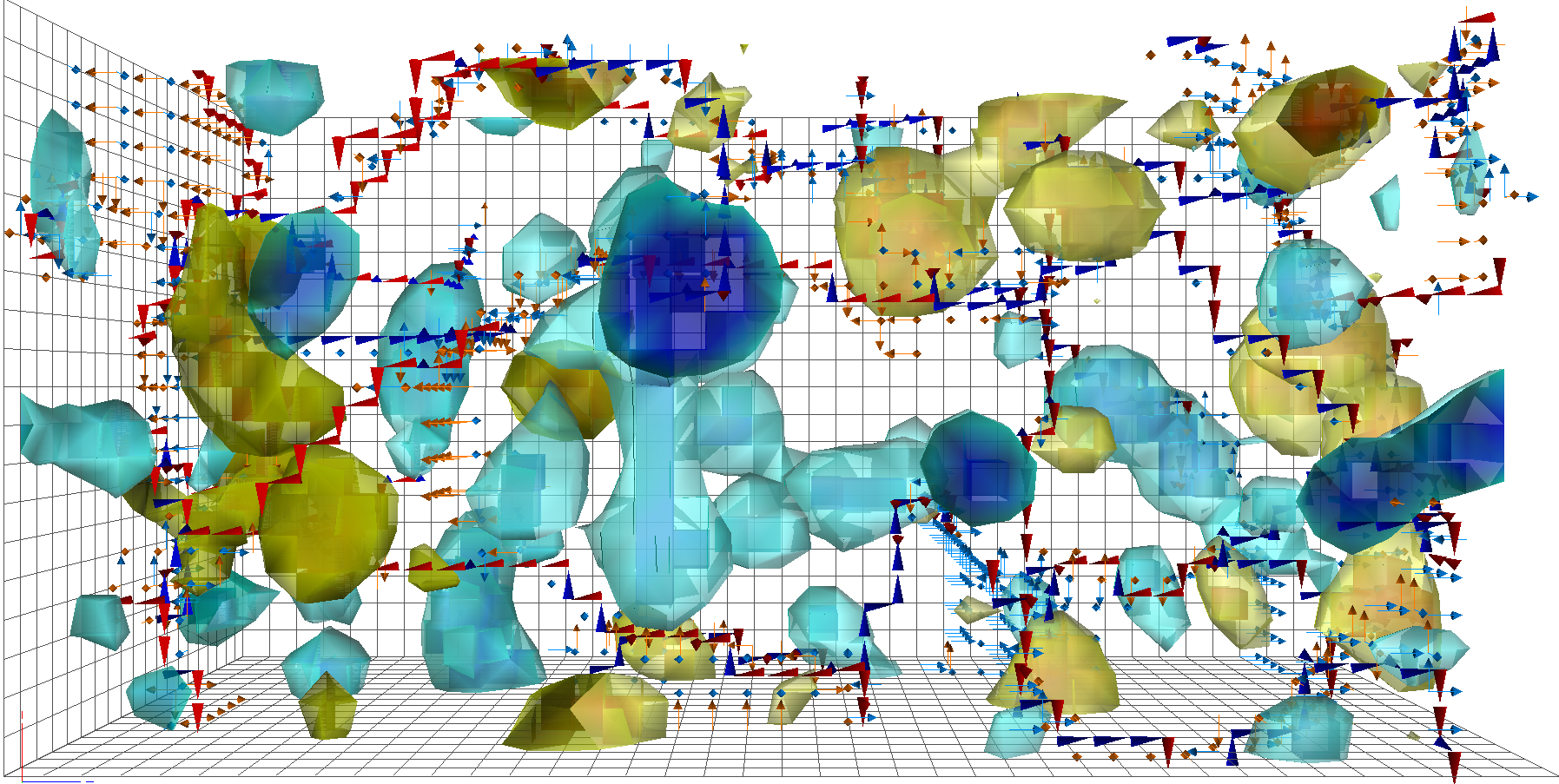}
  \caption{The vortex structure and topological charge present after five sweeps of
    over-improved stout-link smearing, preconditioned with maximal centre gauge
    ($Z^\text{PS}_\mu(x)$ and $q^\text{PS}(x)$). (\textbf{Interactive})}
  \label{fig:MCG5sw}
\end{figure}

Of particular interest is the fact that the correlation does not improve when the vortex configuration is preconditioned by the same degree of smoothing as the topological charge, as shown in Fig.~\ref{fig:AllCorrelation} (c) and (d). This suggests that the primary cause of the more subtle correlation is the vortex projection rather than the smoothing. In fact, we even observe that the correlation shifts closer further from the ideal value of $1$ when the vortex configuration is obtained following five sweeps of smoothing. This arises because the number of vortex structures is reduced under smearing, as seen in Fig.~\ref{fig:MCG5sw}, but the overlap with topological charge has clearly not improved substantially. As noted earlier in Fig.~\ref{fig:PlaqLinkTopQ_SW8}, under cooling this sparsity of vortices is further amplified, indicating that as the degree of smoothing increases, vortices are increasingly removed from the lattice.

An additional consideration for the observed correlation is the fact that projected vortices do not perfectly correlate with the location of the physical thick vortices. Rather, the projected vortices appear within the thick vortex core, but under different Gribov copies of maximal centre gauge they will be identified at different specific lattice sites~\cite{DelDebbio:1998luz}. This variability can contribute to the more subtle correlation observed in Fig.~\ref{fig:AllCorrelation}.

These findings reinforce the result that whilst centre vortices reproduce many of the salient features of QCD, vortex-only configurations are only subtly correlated with the topological charge density of the configurations from which they are obtained. However, the correlation does exist and is consistent across the ensemble, indicating that centre vortices are connected to the topological charge density of the lattice.

\section{Conclusions}
In this work we have presented a new way to examine the four-dimensional structure of centre vortices on the lattice through the use of 3D visualisation techniques. These visualisations give new insight into the geometry and Euclidean time-evolution of centre vortices, and reveal a prevalence of singular points and branching points in the vortex vacuum. It is especially remarkable how common branching points are in SU(3) gauge theory.

We have also explored the connection between these vortex structures and topological charge density. While demonstrating that the topological charge density obtained on projected vortex configurations is generated by singular points, we discovered an interesting correlation between branching points and topological charge; namely that branching points provide an important mechanism for generating large values of topological charge density.

We explored the connection with topological charge density obtained from the original configurations after varying degrees of smoothing. We deduced that the topological charge density of the gauge fields is significantly affected under centre projection, however the modification maintains a positive correlation with the original topological charge density identified on the lattice.

Future work exploring the nature of the Gribov copy problem in regard to $SU(3)$ vortex
locations is of interest, as is an investigation into methods for identifying thick vortex
objects. Additionally, exploration of the change in vortex structure as the temperature
tends towards the deconfining phase is an exciting area of future research. From this work,
it is clear that visualisations of centre vortices provide valuable information about the
structure of the QCD vacuum and provide an elegant complement to numerical results.

\begin{acknowledgments}
  The authors wish to thank Amalie Trewartha for her contributions to the gauge ensembles
  underlying this investigation. We also thank Ian Curington of Visual Technology Services
  Ltd. for his support of the PDF3D software used in creating the interactive 3D
  visualisations presented herein. This research is supported with supercomputing
  resources provided by the Phoenix HPC service at the University of Adelaide and the
  National Computational Infrastructure (NCI) supported by the Australian Government. This
  research is supported by the Australian Research Council through Grants No. DP190102215,
  No. DP150103164, No. DP190100297, and No. LE190100021.
\end{acknowledgments}

\bibliography{VortexVisualisation_static}
\end{document}